\documentclass[prd,12pt,nofootinbib,showpacs,tightenlines]{revtex4}

\usepackage[frenchb,english]{babel}
\usepackage[latin1]{inputenc}
\usepackage{enumerate}
\usepackage{amsmath}
\usepackage{amsthm}
\usepackage{amssymb}
\usepackage{graphicx}
\usepackage{floatflt}
\usepackage{fancybox}
\usepackage{appendix}
\usepackage{enumitem}

\usepackage{hyperref}
\usepackage[b]{esvect}

\usepackage{color}

\renewcommand{\Im}{\textrm{Im}}

\newcommand{\om}{\omega}

\newcommand{\eps}{\epsilon}

\newcommand{\eff}{{\rm eff}}

\renewcommand{\u}{{u}}
\newcommand{\+}{{+}}
\renewcommand{\-}{{-}}
\renewcommand{\v}{{d}}
\newcommand{\hp}{{-u}} 

\newcommand{\omax}{\om_{\rm max}} 

\newcommand{\M}{{\rm max}}

\newcommand{\conj}{{\chi}}

\newcommand{\p}{\partial}

\newcommand{\be} {\begin{equation}}
\newcommand{\ee} {\end{equation}}
\newcommand{\bsub}{\begin{subequations}}
\newcommand{\esub}{\end{subequations}}
\newcommand{\bea}{\begin{eqnarray}}
\newcommand{\eea}{\end{eqnarray}}
\newcommand{\bi} {\begin{itemize}}
\newcommand{\ei} {\end{itemize}}
\newcommand{\ben} {\begin{enumerate}}
\newcommand{\een} {\end{enumerate}}
\newcommand{\bmat} {\begin{pmatrix}}
\newcommand{\emat} {\end{pmatrix}}
\newcommand{\bal} {\begin{aligned}}
\newcommand{\eal} {\end{aligned}}
\newcommand{\btab}{\begin{tabular}}
\newcommand{\etab}{\end{tabular}}

\newcommand{\Sec}[1]{section~\ref{#1}}
\newcommand{\eq}[1]{equation~\eqref{#1}}

\definecolor{myRed}{rgb}{0.8,0.1,0.1}
\definecolor{myGreen}{rgb}{0.7,0,0.8}
\definecolor{myGray}{rgb}{0.6,0.6,0.6}

\begin{document}
\selectlanguage{english}

\title{Low frequency analogue Hawking radiation: \\ The Bogoliubov-de Gennes model}

\author{Antonin Coutant}
\email{antonin.coutant@nottingham.ac.uk}
\affiliation{School of Mathematical Sciences, University of Nottingham, University Park, Nottingham, NG7 2RD, United Kingdom \\
Centre for the Mathematics and Theoretical Physics of Quantum Non-Equilibrium Systems,
University of Nottingham, Nottingham NG7 2RD, United Kingdom}
\author{Silke Weinfurtner}
\email{silke.weinfurtner@nottingham.ac.uk}
\affiliation{School of Mathematical Sciences, University of Nottingham, University Park, Nottingham, NG7 2RD, United Kingdom \\
Centre for the Mathematics and Theoretical Physics of Quantum Non-Equilibrium Systems,
University of Nottingham, Nottingham NG7 2RD, United Kingdom}

\begin{abstract}
We analytically study the low-frequency properties of the analogue Hawking effect in Bose-Einstein condensates. We show that in one-dimensional flows displaying an analogue horizon, the Hawking effect is dominant in the low-frequency regime. This happens despite non vanishing greybody factors, that is, the coupling of the Hawking mode and its partner to the mode propagating with the flow. To show this, we obtained analytical expressions for the scattering coefficients, in general flows and taking into account the full Bogoliubov dispersion relation. We discuss the obtained expressions for the greybody factors. In particular, we show that they can be significantly decreased if the flow obeys a conformal coupling condition. We argue that in the presence of a small but non-zero temperature, reducing greybody factors greatly facilitates the observation of entanglement, that is, establishing that the state of the Hawking mode and its partner is non-separable.
\end{abstract}


\maketitle


\section{Introduction}
In 1981, Unruh discovered an analogy between the propagation of sound in a flowing fluid, and that of waves around a black hole~\cite{Unruh81}. He immediately realized that this analogy could be used to build an experiment to detect, and study the mechanism of the Hawking radiation~\cite{Hawking75}, which predicts that black holes spontaneously emit a flux of thermal radiation. In the recent years, this analogy has received an increasing interest from various experimental groups, and several signatures of the effect has been observed in various experimental setups~\cite{Weinfurtner10,Belgiorno10,Euve15,Steinhauer15,Torres16}.

In this work, we investigate the properties of the analogue Hawking effect in Bose-Einstein condensates. The differences between the spontaneous emission of phonons in a Bose-Einstein condensate and the prediction of Hawking have two origins. First, when the wavelength decreases, the dispersion relation is no longer linear~\cite{Jacobson91,Macher09b}, an assumption the gravitational analogy is based upon. Second, modes that propagate along with the flow are not responsible for the Hawking effect, but they affect the observables by coupling to the Hawking mode and its partner. In black hole physics, they give rise to the so-called greybody factors~\cite{Page76}, which can significantly effect the emitted spectrum. A particular consequence of these greybody factors is that the low-frequency emission spectrum of a black hole is suppressed, because low-frequency modes do not have enough energy to escape from the gravitational potential. On the contrary, in one-dimensional acoustic flows, it was discovered that the spectrum of the Hawking effect is dominant at low frequencies~\cite{Macher09b,Mayoral10,Coutant13,Anderson14,Anderson15,Fabbri16}. 

In this work, we study the low-frequency properties of the Hawking effect, and demonstrate that the spectrum increases at low frequencies in the most general case. We consider both greybody factors and the effect of dispersion (not considered in~\cite{Anderson14,Anderson15,Fabbri16}). For this, we use a generalization of the matched asymptotic expansion method~\cite{LandauV3,Brito15}, that allows us to fully characterize the scattering at low frequencies.  We discuss this scattering in the most general class of flows that can be obtained in condensates. This allows us to characterize greybody factors, but also to point out how to minimize them, and hence recover predictions very close to the ideal Hawking case.

In a companion work~\cite{Coutant17}, we used the same method to characterize the low-frequency effects of dispersion. We worked in a simpler model, the linear Korteweg-de Vries model, which neglects the presence of the downstream mode. In this work on the contrary, we consider the full Bogoliubov-de Gennes equation, which describes the propagation of sound waves in a Bose-Einstein condensate. We shall not only discuss the emitted spectrum, but also how the entanglement between the Hawking mode and its partner is affected by the coupling to the downstream mode.

In section~II, we briefly review the Bogoliubov approximation of the excitations of a condensate, and the definition of the scattering matrix on a varying flow. In section~III, we present the method and obtain the general behavior of the coefficients of the scattering matrix. In section~IV, we discuss the various consequences of greybody factors, and how to control them. The conclusion are complemented by the study of two exactly solvable examples. We work in units where $\hbar = k_B = 1$.

\section{The Bogoliubov-de Gennes model}
\subsection{Wave equation}
\begin{figure}[!ht]
\begin{center}
\includegraphics[width=0.49\columnwidth]{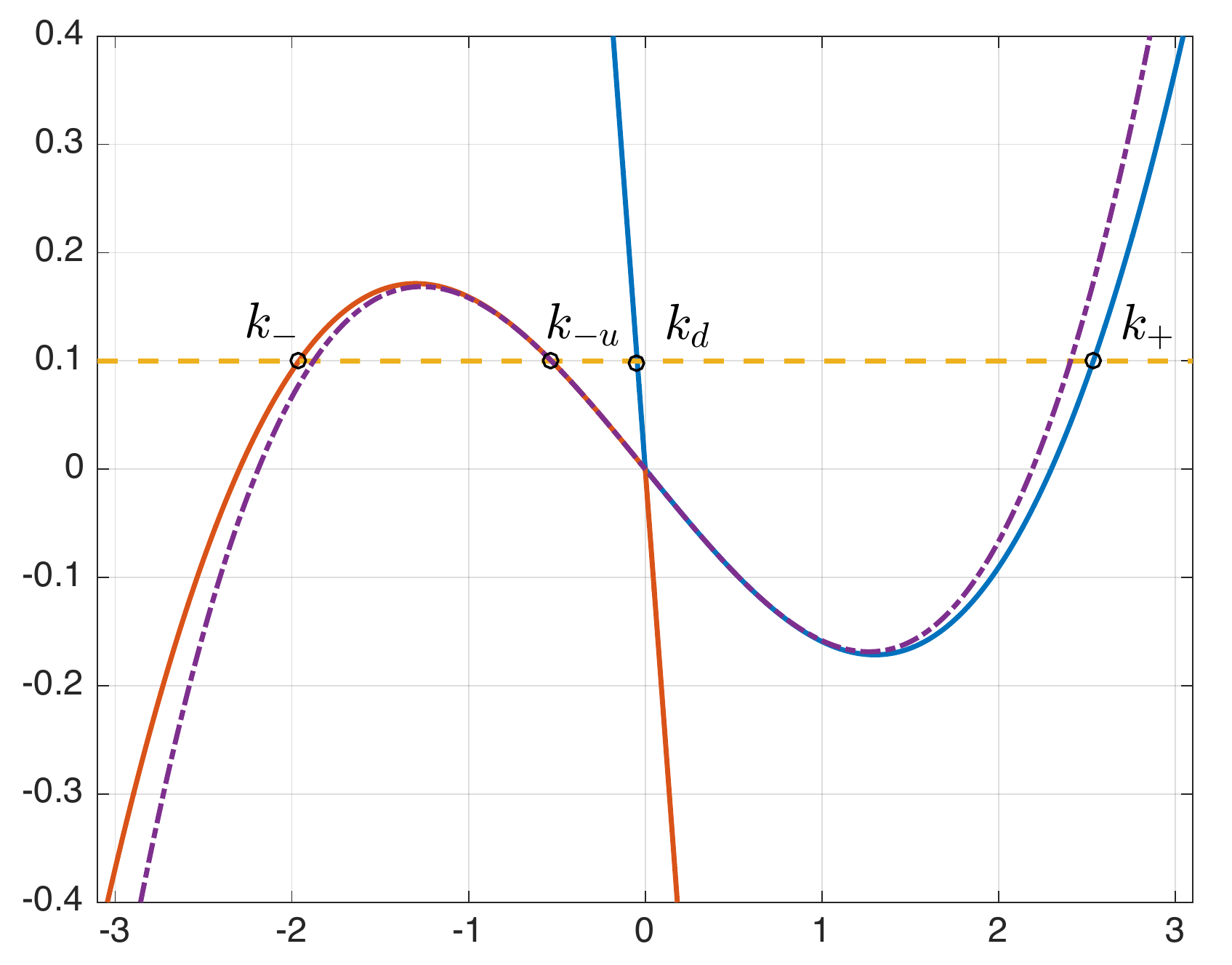}
\includegraphics[width=0.49\columnwidth]{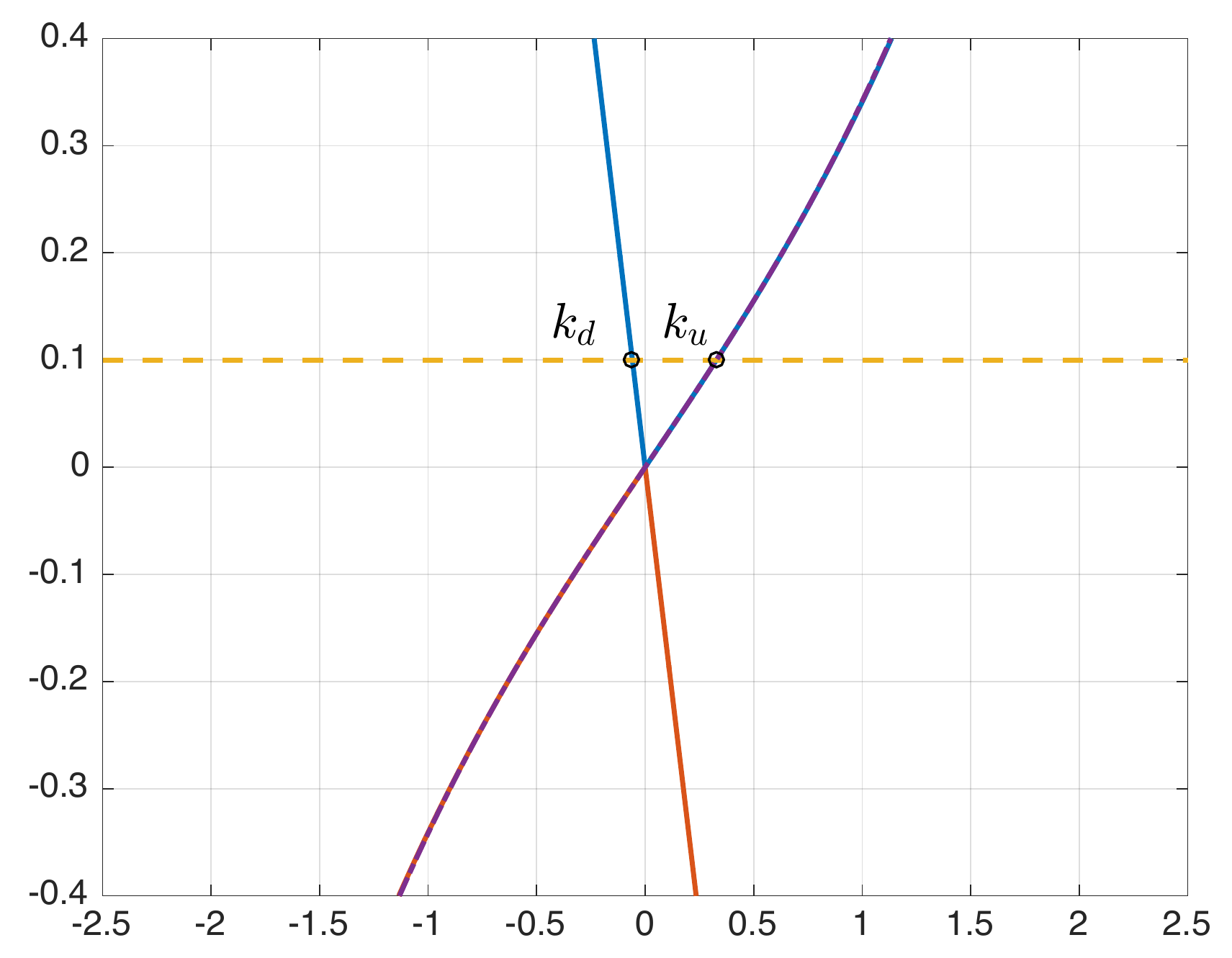}
\end{center}
\caption{$\om$ as a function of $k$ from the dispersion relation \eqref{HJ}, in units where $c=1$ and $\xi = 1$. On the left side, in a supersonic flow ($v=-1.2$), on the right side, in a subsonic one ($v=-0.7$). The dashed line shows the number of solutions for $\om = 0.1$. The value of $\om$ such that $k_- = k_\hp$ defines $\omax$. To compare with the work of~\cite{Coutant17}, the dotted-dash line represents the Korteweg-de Vries approximation, that is, $\om = (c+v)k + k^3/(8m^2c)$.
}
\label{HJ_Fig}
\end{figure}

In this section, we briefly present the problem of scattering of density waves in transonic flows. We refer to the literature for a more detailed account and the precise relation to black hole physics~\cite{Garay00,Balbinot08,Macher09b,Larre12}. We consider a gas made of $N$ identical spinless bosons of mass $m$ with a point-like interaction. In the one-dimensional regime~\cite{Petrov00,Pitaevskii} (when transverse trapping frequencies are higher than typical energy scales along the preferred direction)~\footnote{The first changes for the scattering introduced by a breakdown of the one-dimensional approximation will be the coupling to modes with a mass gap. It is noticeable that such modes have different low-frequency properties, as shown in~\cite{Coutant12}.}, the bosonic field operator $\hat \Psi$ evolves according to
\be \label{N_body}
i \p_t \hat \Psi = - \frac{\p_x^2 \hat \Psi}{2m} + V_{\rm ext}(x) \hat \Psi + g(x) \hat \Psi^\dagger \hat \Psi \hat \Psi.
\ee
$V_{\rm ext}$ is the external potential, and $g$ the (effective one-dimensional) coupling constant of inter-particle interaction. To consider the most general case, we allow both to depend on the position. However, we restrict ourselves to stationary backgrounds; hence, $V_{\rm ext}$ and $g$ are independent of time $t$. We assume the temperature to be low enough for the system to be in the quasi-condensate regime~\cite{Pitaevskii}. In this regime, the condensate wave function $\langle \hat \Psi \rangle = \psi_0 e^{- i \mu t}$ obeys the stationary Gross-Pitaevski equation, i.e.
\be \label{GP_eq}
\mu \psi_0 = - \frac{\p_x^2 \psi_0}{2m} + V_{\rm ext}(x) \psi_0 + g(x) |\psi_0|^2 \psi_0,
\ee
where $\mu$ is the chemical potential. The condensate wave function defines the density $\rho$ and velocity $v$ of the flow
\be
\psi_0(x) = \sqrt{\rho(x)} e^{i m \int v(x') dx'}.
\ee
Perturbations around this condensate are described by a phononic field operator $\hat \phi$ defined by
\be
\hat \Psi = \psi_0 e^{- i \mu t} \left({\bf 1} + \hat \phi\right).
\ee
The field operator $\hat \phi$ can be decomposed into a superposition of stationary modes
\be \label{field_decomp}
\hat \phi(t,x) = \sum \int \left( \hat a_\om \phi_\om (x) e^{-i \om t} + \hat a_\om^\dagger \varphi_\om^*(x) e^{i \om t}\right) \frac{d\om}{\sqrt{2\pi}} ,
\ee
where the sum runs over the different modes of the same frequency $\om$. Linearizing equation \eqref{N_body} in $\hat \phi$, we obtain the field equations for the modes $\phi_\om$ and $\varphi_\om$ as
\bsub \label{field_eq} \bea
\left(\om + i v \p_x \right) \phi_\om &=& \frac{-1}{2m \rho} \p_x \rho \p_x \phi_\om + g \rho \left(\phi_\om + \varphi_\om \right) , \\
\left(\om + i v \p_x \right) \varphi_\om &=& \frac1{2m \rho} \p_x \rho \p_x \varphi_\om - g \rho \left(\phi_\om + \varphi_\om \right) .
\eea \esub
To relate this equation to physical observables, we will use phase fluctuations $\theta_\om$ and relative density fluctuations $n_\om = \delta \rho/(2\rho)$ instead of the field modes $\phi_\om$, and $\varphi_\om$. It turns out that the mode \eq{field_eq} is also easier to solve using $\theta_\om$ and $n_\om$. The two are directly related of field modes by
\bsub \bea
\theta_\om &=& \frac1{2i}(\phi_\om - \varphi_\om), \\
n_\om &=& \frac12 (\phi_\om + \varphi_\om) .
\eea \esub
They obey the set of equations
\bsub \label{mode_eq} \bea
\left(-i \om + v \p_x \right) \theta_\om &=& \frac1{2m \rho} \p_x \rho \p_x n_\om - 2 g \rho n_\om , \label{mode_eq1} \\
\left(-i \om + v \p_x \right) n_\om &=& -\frac1{2m \rho} \p_x \rho \p_x \theta_\om . \label{mode_eq2}
\eea \esub
We start by considering a homogeneous condensate flowing to the left ($v<0$). Since the background quantities are independent of $x$, solutions of \eqref{mode_eq} are given by plane waves
\bsub \label{plane_waves} \bea
\theta_\om &=& U_k e^{- i \om t + i k x} , \\
n_\om &=& V_k e^{- i \om t + i k x}.
\eea \esub
The frequency $\om$ and the wave number $k$ are related by the dispersion relation
\be \label{HJ}
(\om - v k)^2 = c^2 k^2 + \frac{k^4}{4m^2}.
\ee
In this equation, we have defined the speed of sound $c^2 = g \rho/m$, which gives the propagation speed of long wavelength waves. Using this velocity $c$, one can build a characteristic length
\be
\xi = \frac1{2 m c}.
\ee
$\xi$ is the healing length. In \eq{HJ}, it determines the length below which dispersive effects become significant. In the next section, we will solve explicitly this equation for low frequencies. We first classify the various roots of equation \eqref{HJ}, using its graphical resolution in Fig.~\ref{HJ_Fig}. When the flow is subsonic $|v| < c$, there are always two solutions, one moving upstream (noted $k_\u$), one moving downstream (noted $k_\v$). When the flow is supersonic, and below a certain frequency $0 < \om < \om_\M$, there are 4 solutions. Two of them correspond to the usual right and left movers, but the one going against the flow propagates too slowly, and hence is dragged by the flow. The peculiarity is that this root corresponds to a mode of negative energy (or norm, see below). In addition, we have two extra solutions. These solutions are allowed by the non-zero flow and by dispersion, i.e. they disappear in the limit $\xi \to 0$. For this reason, we shall refer to them as the dispersive roots. One of them has a positive energy, while the other have a negative energy. We will denote them $k_\+$ and $k_\-$, where the subscript refers to the sign of the energy.
In addition to the dispersion relation, the mode \eq{mode_eq} gives us a linear relation between the amplitudes $U_k$ and $V_k$ defined in \eq{plane_waves}. Using \eqref{mode_eq2}, we obtain
\be
V_k = i \frac{k^2}{2m (\om - v k)} U_k.
\ee
This means that the phase operator $\hat \theta = (\hat \phi - \hat \phi^\dagger)/2i$ decomposes into eigen-modes in the following manner:
\be
\hat \theta(t,x) = \sum_j \int \left[ \hat b_j U_{k_j} e^{- i \om_j t + i k_j x} \right] \frac{d\om}{\sqrt{2\pi}} + h.c. ,
\ee
where the index $j$ varies over the solution of the dispersion relation at fixed $\omega$. The density fluctuations operator $\hat n = (\hat \phi + \hat \phi^\dagger)/2$ has a the same decomposition, with the amplitudes $V_k$ instead of $U_k$. The canonical commutation relation $[\hat \phi, \hat \phi^\dagger] = \rho \delta(x-x')$ now gives us the commutation relations between the operators $\hat b_j$'s. Then, to identify them as creation and annihilation operators, satisfying the canonical commutation relation
\be
[a_{\om', j'}, a_{\om, j}^\dagger] = \delta_{j j'} \delta(\om - \om'),
\ee
we normalize the amplitudes such that
\be \label{normaliz_eq}
\Im(V_k U_k^*) = \frac{\pm 1}{4 \rho |v_g(k)|} ,
\ee
(see e.g.~\cite{Macher09b} for more details). The sign in \eq{normaliz_eq}, is the sign of the norm of the corresponding mode. It indicates if that mode is associated with a creation or annihilation operator~\cite{Dalfovo99,Macher09b}. In \eq{normaliz_eq}, we have also defined $v_g$, the group velocity associated with the mode of wave number $k$, i.e. $v_g = (\p_k \om)^{-1}$. For simplicity, we can choose the phase reference such that $U_k$ is real. Then the normalization is given by
\bsub \label{norms} \bea
U_k &=& \sqrt{\left|\frac{2 m (\om - v k)}{\rho k^2 v_g}\right|} , \\
V_k &=& \pm i \sqrt{\left|\frac{k^2}{8 m \rho v_g (\om - v k)}\right|}.
\eea \esub


\subsection{Varying background and $S$-matrix}
\label{Bckgr_Sec}
We now assume that the flow accelerates, or decelerates, over a region of length scale $L$, and centered around $x=0$ for commodity. When $x \ll -L$ or $x \gg L$, the flow is constant, that is
\bsub \bea \label{Vas}
V_{\rm ext}(x) &\to& V_{r/l} , \\
g(x) &\to& g_{r/l} , \\
\rho(x) &\to& \rho_{r/l} , \\
v(x) &\to& -v_{r/l} , \\
c(x) &\to& c_{r/l} ,
\eea \esub
where the subscript $l$ (resp. $r$) is for $x \to -\infty$ (resp. $+\infty$). Notice also that the condensate flows to the left, hence $v_{r/l} > 0$. The 5 background functions in \eqref{Vas} are not independent, since the Gross-Pitaevski \eq{GP_eq} must be satisfied. The first relation is the continuity equation, which, for stationary unidimensional flows, gives
\be \label{cont_eq}
\rho v = {\rm const.}
\ee
The second equation obtained from the Gross-Pitaevski equation~\eqref{GP_eq} gives a relation between $V_{\rm ext}(x)$ and the density $\rho$. In addition, the definition of the speed of sound gives us $c^2(x) = g(x) \rho(x)/m$. Using these three relations, the background is fully characterized by two independent background functions out of \eq{Vas}. In other words, by choosing the external potential $V_{\rm ext}(x)$ and the coupling constant $g(x)$ in \eqref{GP_eq}, one can impose a certain profile for $\rho$, $v$, and $c$. In an experiment, these functions might be delicate to control with the necessary precision. Possibilities to have a varying coupling constant $g$ are to adiabatically change the transverse trapping frequency~\cite{Pavloff02}, or to exploit a Feschbach resonance~\cite{Pitaevskii,Cornish00}. In this work, instead of $V_{\rm ext}$ and $g$, we shall use $v(x)$ and $c(x)$ to fully determine the propagation equation of waves in the condensate through \eq{mode_eq}, and assume generic form for their profiles.

\begin{figure}[!ht]
\begin{center}
\includegraphics[width=0.8\columnwidth]{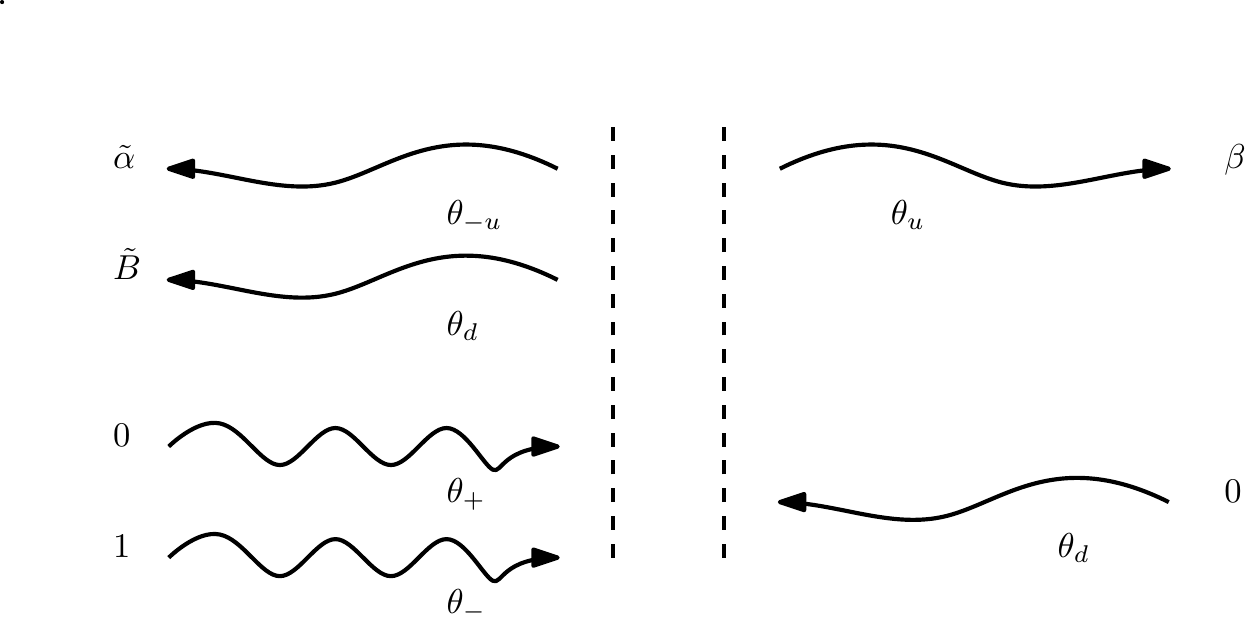}
\end{center}
\caption{Schematic representation of the scattering problem. We have represented the amplitudes the asymptotic behavior of $\theta_{-}^{\rm in}$ on a black hole flow (see \eq{Smat}).
}
\label{KU_Fig}
\end{figure}

On both asymptotic sides $|x| \gg L$, the solutions of \eqref{mode_eq} are given by superpositions of plane waves. From the earlier discussion of the dispersion relation \eqref{HJ}, there are 2 asymptotic modes on the subsonic side and 4 on the supersonic one (since we are interested in low frequencies, we always assume $\om < \om_\M$). As read from their group velocity, 3 of them propagate towards the transition region, while 3 of them propagate away from it. The linear relation between modes going in and modes coming out defines the $S$-matrix:
\be \label{Smat}
\bmat \theta_\+^{\rm in} \\ \theta_\-^{\rm in} \\ \theta_\v^{\rm in} \emat = S \cdot \bmat \theta_\u^{\rm out} \\ \theta_\hp^{\rm out} \\ \theta_\v^{\rm out} \emat =
\bmat \alpha & \tilde \beta & \tilde R \\ \beta & \tilde \alpha & \tilde B \\ R & B & \tilde T \emat \cdot \bmat \theta_\u^{\rm out} \\ \theta_\hp^{\rm out} \\ \theta_\v^{\rm out} \emat.
\ee
This is illustrated in Fig.~\ref{KU_Fig}. Because we work with normalized modes, the conservation of the norm (or energy) implies that $S \in U(2,1)$. However, as we shall see, at low frequencies, the dispersive modes $\theta_\pm$ tend to have the same amplitude, due to the fact that $|\alpha| \sim |\beta|$ and $|\tilde \alpha| \sim |\tilde \beta|$ (see equations \eqref{thetaU} and \eqref{thetaHP} below). Therefore it will be difficult to build the scattering modes $\theta_\pm^{\rm in}$. To circumvent the problem, we shall construct \emph{out} modes instead. We then obtain the scattering coefficients by inverting the $S$-matrix. Since $S \in U(2,1)$, this inversion is fairly simple, and we have
\be \label{Sdag}
S^{-1} = \bmat \alpha^* & - \beta^* & R^* \\ -\tilde \beta^* & \tilde \alpha^* & - B^* \\ \tilde R^* & - \tilde B^* & \tilde T^* \emat.
\ee
The minus signs follows from the fact that $S \in U(2,1)$. Moreover, this property of $S$ also gives several relations between the coefficients, such as
\bsub \bea
|\alpha|^2 - |\beta|^2 + |R|^2 &=& 1, \\
|\tilde \alpha|^2 - |\tilde \beta|^2 - |B|^2 &=& 1, \\
|\tilde R|^2 - |\tilde B|^2 + |\tilde T|^2 &=& 1.
\eea \esub
Both \emph{in} and \emph{out} basis allow for a decomposition of the field operators, i.e.
\bsub \bea
\hat \theta(t,x) &=& \int_0^{\omax} \left[\hat a_\u^{\rm in} \theta_\u^{\rm in} + (\hat a_\hp^{\rm in})^\dagger \theta_\hp^{\rm in} + \hat a_\v^{\rm in} \theta_\u^{\rm in}\right] e^{-i \om t} d\om + h.c. , \\
&=&  \int_0^{\omax} \left[\hat a_\+^{\rm out} \theta_\+^{\rm out} + (\hat a_\-^{\rm out})^\dagger \theta_\-^{\rm out} + \hat a_\v^{\rm out} \theta_\u^{\rm out} \right] e^{-i \om t} d\om + h.c.
\eea \esub
(Since we focus in this work on low frequencies, we have omitted frequencies $\om > \omax$. In that case the scattering becomes elastic and $2\times 2$~\cite{Macher09b}.) The two possible decompositions of this operator leads to a linear relation between the $\hat a^{\rm out}$'s and $\hat a^{\rm in}$'s, inherited from the $S$-matrix, which reads
\be \label{aaS}
\bmat \hat a_\u^{\rm out} \\ (\hat a_\hp^{\rm out})^\dagger \\ \hat a_\v^{\rm out} \emat = \bmat \alpha & \beta & R \\ \tilde \beta & \tilde \alpha & B \\ \tilde R & \tilde B & \tilde T \emat \cdot \bmat \hat a_\+^{\rm in} \\ (\hat a_\-^{\rm in})^\dagger \\ \hat a_\v^{\rm in} \emat.
\ee
When incoming modes are in their ground states, because the scattering mixes positive and negative norm modes, there is a spontaneous emission of phonons. This is directly obtained from \eq{aaS}. For instance, the flux of $\u$-modes is given by
\be
n_\u \equiv \langle 0_{\rm in} | (\hat a_\u^{\rm out})^\dagger \hat a_\u^{\rm out} | 0_{\rm in} \rangle = |\beta|^2.
\ee
And similarly for the emission of the other modes $n_\hp$, $n_\v$. When neglecting dispersion and the mode $\theta_\v$, the emission spectrum of out-going modes follows a Planck distribution:
\be \label{HR}
|\beta|^2 = \frac1{e^{\om/T_H} - 1},
\ee
where the Hawking temperature is given by the gradient of the flow at the point $\mathcal H$ where $|v| = c$ (equivalently, by the surface gravity of the analogue horizon), that is
\be \label{THR}
T_H = \frac{\kappa}{2\pi} = \frac1{2\pi} \p_x(c+v)_{\mathcal H}.
\ee
It has been shown in several works~\cite{Unruh95,Corley96,Brout95,Himemoto00,Unruh04,Coutant11,Coutant14b,Philbin16} that \eqref{HR} is maintained when $L \gg \xi$, and when $\theta_\v$ is neglected. In this work, we relax these two assumptions, and study in depth the low-frequency behavior of the coefficients. In the Hawking regime, the low-frequency behavior is simple, as it follows from \eqref{HR} that
\be \label{HR0}
|\beta|^2 \sim \frac{T_H}{\om}.
\ee
This result is in general affected by the coupling to the downstream mode $\theta_\v$. The dressing of the flux $n_\u$ by the downstream mode is referred to as greybody factors in the black hole literature~\cite{Page76}. In the rest of the paper, we shall focus particularly on the coefficients $R$, and $B$, which are the generalization of the greybody factors~\footnote{In the black hole literature~\cite{Page76}, only $R$ is considered as a greybody factor, because the Hawking flux is given by \eqref{HR} times $1-|R|^2$. On the contrary, $B$ changes the flux of the partner, which is emitted inside the hole. In analogue flows however, the partner flux is accessible, and provides valuable information, such as the correlations with the Hawking mode~\cite{Balbinot08,Parentani10,Schutzhold10} that we shall discuss in section~\ref{Entang_Sec}.} for an acoustic black hole in a Bose-Einstein condensate. As we shall see in section~\ref{Coefs_Sec}, these coefficients not only dress the emitted spectrum, but they also alter the amount of entanglement between the Hawking mode $\theta_\u$ and its partner $\theta_\hp$. Lastly, we point out that our conclusions will equally apply to white hole flows (time-reverse of black holes), since the corresponding $S$-matrix is simply $S^{-1}$ (see~\cite{Coutant11} or the appendix D of~\cite{Macher09b} for more details).


\section{Low-frequency scattering}
\subsection{Method and mode basis}
In this paper, we will analyze the properties of the $S$-matrix in the limit of small frequencies. For this, we divide the problem into two regions. In the far region, for $|x| \gg L$, the background is homogeneous, and solutions are a superposition of plane waves. In the near region, for $|x| \ll c/\om$ (where $c$ is here the typical value of $c(x)$ along the flow), $\om$ can be neglected in the mode \eq{mode_eq}, which becomes easier to solve. Then, we identify the large $x$ behavior of the solutions obtained for $\om \to 0$ to the low-frequency expansion of the mode basis in a matching region such that $L \ll |x| \ll c/\om$. Therefore, this procedure is valid under the condition that this matching region is large enough. This gives us the validity regime of the low-frequency scattering coefficients we will obtain, namely
\be \label{valid_cond}
\frac{\om L}c \ll 1.
\ee
This method is is known as \emph{matched asymptotic expansion}. It is wildly used for second-order differential equations, in quantum mechanics~\cite{LandauV3}, or black hole physics~\cite{Starobinskii73,Page76}, and has also been used to analyse greybody factors in acoustic metrics~\cite{Anderson14,Anderson15}. What we present here is a generalization of this method to \eq{mode_eq} (which can be recast as a fourth-order differential equation~\cite{Macher09b}), in order to take into account dispersive effects. The first step is to build the mode basis at low frequencies. To enlighten the notations along the paper, we first introduce the characteristic wave numbers on the left and right sides as
\bsub \label{q_def} \bea
q_l &=& 2 m \sqrt{v_l^2 - c_l^2} , \\
q_r &=& 2 m \sqrt{v_r^2 - c_r^2} .
\eea \esub
We now solve the dispersion relation \eqref{HJ} for low $\om$, on both asymptotic sides. For $x \gg L$, they read
\bsub \bea
k_\u &=& \frac{\om}{c_r - v_r} + O(\om^2) , \\
k_\v &=& -\frac{\om}{c_r + v_r} + O(\om^2) ,
\eea \esub
while for $x \ll -L$, we have
\bsub \bea
k_\+ &=& q_l + \frac{\om v_l}{v_l^2 - c_l^2} + O(\om^2), \\
k_\- &=& - q_l + \frac{\om v_l}{v_l^2 - c_l^2} + O(\om^2), \\
k_\hp &=& -\frac{\om}{v_l - c_l} + O(\om^2), \\
k_\v &=& -\frac{\om}{c_l + v_l} + O(\om^2) .
\eea \esub
Note that we have kept the subleading term in $O(\om)$ for $k_\pm$ in order to obtain their group velocity. We can now build the mode basis on each side by using the above roots of the dispersion relation on plane wave solutions given by \eqref{plane_waves}, and \eqref{norms}. We then evaluate them for $|x| \ll c/\om$. On the right side we obtain
\bsub \label{basis_r} \bea
\theta_\u &=& \sqrt{\frac{m c_r}{2\om \rho_r}} e^{i k_\u x} \underset{|x \om| \gg c}\sim \sqrt{\frac{m c_r}{2\om \rho_r}} + i\frac{\sqrt{m c_r \om}}{(c_r-v_r)\sqrt{2\rho_r}}x , \\
\theta_\v &=& \sqrt{\frac{m c_r}{2\om \rho_r}} e^{i k_\v x} \underset{|x \om| \gg c}\sim \sqrt{\frac{m c_r}{2\om \rho_r}} - i\frac{\sqrt{m c_r \om}}{(c_r+v_r)\sqrt{2\rho_r}}x ,
\eea \esub
and on the left we have
\bsub \label{basis_l} \bea
\theta_\+ &=& U_Z e^{i q_l x}, \\
\theta_\- &=& U_Z e^{- i q_l x}, \\
\theta_\hp &=& \sqrt{\frac{m c_l}{2\om \rho_l}} e^{i k_\hp x} \underset{|x \om| \gg c}\sim \sqrt{\frac{m c_l}{2\om \rho_l}} - i\frac{\sqrt{m c_l \om}}{(v_l - c_l)\sqrt{2 \rho_l}}x , \\
\theta_\v &=& \sqrt{\frac{m c_l}{2\om \rho_l}} e^{i k_\v x} \underset{|x \om| \gg c}\sim \sqrt{\frac{m c_l}{2\om \rho_l}} - i\frac{\sqrt{m c_l \om}}{(c_l+v_l)\sqrt{2\rho_l}}x .
\eea \esub
To simplify, we called $U_Z$ the common normalization of the dispersive modes, which reads
\be
U_Z = \sqrt{\frac{v_l^2}{4\rho_l (v_l^2 - c_l^2)^{3/2}}}.
\ee
In the next section, we will solve the wave equation at low frequencies, and then obtain their asymptotic behavior for $|x| \gg L$. As we shall see, the large $x$ asymptotics of these solutions are a superposition of constant and linear terms in $x$, and oscillating exponentials. The existence of a matching region will allow us to identify these asymptotics to the different plane wave modes obtained in equations \eqref{basis_r} and \eqref{basis_l}. Notice that it is necessary to keep the terms linear in $x$ in the low-frequency expansion of $\theta_\u$, $\theta_\hp$ and $\theta_\v$, as they allow us to distinguish the two different long wavelength modes.

\subsection{General results}
\label{Gen_results_Sec}
We start by setting $\om = 0$ in \eq{mode_eq}. Then, using $\rho v = {\rm const}$, we integrate \eqref{mode_eq2} into
\be \label{nTOtheta0}
n_\om = -\frac1{2 m v} \p_x \theta_\om + a.
\ee
By changing the constant of integration $a$, one obtains different possible modes. We start by setting $a=0$, and will later on consider modes for $a \neq 0$. We then plug the above relation in \eqref{mode_eq1}, and get
\be
4m^2 \rho^3 (c^2 - v^2) \p_x \theta_\om = \rho \p_x \rho \p_x \rho \p_x \theta_\om.
\ee
We define an auxiliary field $\conj$ as
\be
\conj = \rho \p_x \theta_\om.
\ee
The field $\conj$ then obeys the second-order equation
\be \label{Schro_eq}
4m^2 (c^2 - v^2) \conj = \frac1\rho \p_x \rho \p_x \conj.
\ee
This is already a notable result. Namely, the low-frequency behavior of phonon scattering is entirely determined by a Schrödinger-like equation in a potential, for which many tools and solvable examples are know. To start, we will show how the general asymptotic properties of this equation gives the frequency dependence of the scattering coefficients. Since we consider a transonic flow, we have $c_l < v_l$, and $v_r < c_r$. Using the definitions of \eqref{q_def}, the solutions of \eqref{Schro_eq} are asymptotically exponentials with rate $q_{l/r}$. On the left side, the exponentials oscillate ($c_l^2 < v_l^2$), while on the right side ($c_r^2 > v_r^2$) they grow or decay. Explicitly, the asymptotic expansion of $\conj$ has the form
\bsub \label{chi_gen} \bea
\conj &\underset{-\infty}{\sim}& A_2 e^{i q_l x} + A_3 e^{-i q_l x}, \\
&\underset{+\infty}{\sim}& A_\uparrow e^{q_r x} + A_\downarrow e^{- q_l x}.
\eea \esub
We underline that the coefficients $A$'s are independent of $\om$, since they are obtained from \eq{Schro_eq}, which do not contain $\om$. Since all scattering states are spatially bounded, we first set $A_\uparrow = 0$. Then, we integrate $\conj$ and obtain the asymptotic behavior of $\theta$:
\bsub \label{1order_uin} \bea
\theta^{1} &\underset{-\infty}{\sim}& A_2 \frac{e^{i q_l x}}{i q_l \rho_l} + A_3 \frac{e^{-i q_l x}}{-i q_l \rho_l} , \\
&\underset{+\infty}{\sim}& A_1.
\eea \esub
Here we have fixed the integration constant to vanish on the left side. This implies that $A_1$ is given by the integral~\footnote{More precisely, $\conj$ decays exponentially for $x \to +\infty$, but oscillates near $-\infty$. Hence, the integral must be regularized, as the limit $\eps \to 0$ of $\int e^{\eps x} \conj / \rho dx$. In the rest of the paper, we use the same regularization prescription for similar integrals. \label{regul_ftn}}
\be \label{A1}
A_1 = \int_{-\infty}^{+\infty} \frac{\conj(x')}{\rho(x')} dx'.
\ee
$\theta_1$ is one particular solution of the mode \eq{mode_eq} at $\om=0$. The first mode we want to obtain is the long wavelength mode coming out to the right, that is $\theta_\u^{\rm out}$. Using the inverse $S$-matrix in \eqref{Sdag}, and the form of the mode basis for $L \ll x \ll 1/\om$, we have
\bsub \label{uINmode} \bea
\theta_\u^{\rm out} &\underset{-\infty}{\sim}& \alpha^* U_Z e^{i q_l x} - \beta^* U_Z e^{-i q_l x} , \\
&\underset{+\infty}{\sim}& (1 + R^*) \sqrt{\frac{m c_r}{2\om \rho_r}}  + i\frac{\sqrt{m c_r \om}}{(c_r-v_r)\sqrt{2\rho_r}}x - i\frac{R^* \sqrt{m c_r \om}}{(c_r + v_r)\sqrt{2 \rho_r}}x .
\eea \esub
We see that $\theta_1$ is a good candidate, since it is purely oscillating on the left side (no constant or linear term in $x$ on the left side). As previously mentioned, since we have two different long wavelength modes ($\theta_\u$ and $\theta_\v$), we must keep the term in $O(x)$ in the limit $\om \to 0$ to be able to distinguish them. As we see from \eq{uINmode}, this term is higher order in $\om$. Hence, one must take into account the first-order correction in $\om$ from the wave equation, and obtain a mode with the asymptotic behavior
\bsub \label{NLOtheta_as} \bea
\theta_1 &\underset{-\infty}{\sim}& A_2 \frac{e^{i q_l x}}{i q_l \rho_l} + A_3 \frac{e^{-i q_l x}}{-i q_l \rho_l} , \\
&\underset{+\infty}{\sim}& A_1 - i \om A_4 x,
\eea \esub
where $A_4$ is independent of $\om$. To obtain $A_4$, we first invert the relation between $\theta_\om$ and $n_\om$ in \eqref{mode_eq2} at first order in $\om$. Using the fact that $\rho v$ is constant, we obtain
\be \label{nTOtheta1}
n_\om = - \frac1{2 m v} \p_x \theta_\om - i \om \int_{-\infty}^x \frac{\p_x \theta_\om}{2 m v^2} dx' + O(\om^2).
\ee
The integration constant is chosen so that there is no constant or linear term in $x$ near $-\infty$. We now inject this in \eq{mode_eq1}, and obtain
\be \label{First_om_corr}
\frac{v^2 - c^2}{v} \p_x \theta_\om = i \om \theta_\om + i \om c^2 \int_{-\infty}^x \frac{\p_x \theta_\om}{v^2} dx' + \frac1{2m \rho} \p_x \rho \p_x n_\om + O(\om^2).
\ee
We now extract $A_4$ by evaluating this for $x \to +\infty$. Using the asymptotic \eqref{NLOtheta_as} and the relation \eqref{nTOtheta1}, we show that $\p_x n_\om$ is $O(\om^2)$. Hence, identifying all terms of order $O(\om)$, we obtain
\be \label{A4}
\frac{(v_r^2-c_r^2)}{v_r} A_4 = A_1 + c_r^2 \int_{-\infty}^{+\infty} \frac{\p_x \theta_{\om = 0}}{v^2} dx.
\ee
To rewrite this equation in a simpler form, we define an ``effective velocity'' as
\be \label{veff_def}
\frac1{v_\eff^2} = \frac1{A_1} \int_{-\infty}^{+\infty} \frac{\p_x \theta_{\om = 0}}{v^2} dx.
\ee
It is easy to see that if $v(x)$ is constant, we simply have $v_\eff = |v|$ ($v_\eff$ is also independent of a choice of normalization and phase reference of $\theta_\om$). Hence, $v_\eff$ can be interpreted as some averaging of the background flow $v(x)$ by the mode. Its interest is that it allows for a compact writing of the grebody factors. The last step to obtain the scattering coefficients is to identify $\mathcal N \theta_1$ to $\theta_\u^{\rm in}$, where $\mathcal N$ is an overall normalizing constant. This gives us 4 equations (2 oscillating components on the left and a $O(1)$ and $O(x)$ on the right), and 4 unknown coefficients: $\alpha$, $\beta$, $R$, and the normalization constant $\mathcal N$. Solving this linear system, we obtain the scattering coefficients:
\bsub \label{thetaU} \bea
R &=& \frac{v_\eff^2 - c_r v_r}{v_\eff^2 + c_r v_r} , \label{R_eq} \\
\alpha &=& i \frac{A_2}{q_l \rho_l U_Z A_1} \sqrt{\frac{m c_r}{2 \om \rho_r}} \times \left(1+R \right) ,\\
\beta &=& i \frac{A_3}{q_l \rho_l U_Z A_1} \sqrt{\frac{m c_r}{2 \om \rho_r}} \times \left(1+R \right).  \label{beta_eq}
\eea \esub
This is our first main result. It gives the low-frequency behavior of the coefficient. We see that at low $\om$, the greybody factor $R$ goes to a constant, i.e. $R = O(\om^0)$, while $\alpha$ and $\beta$ grow like $O(1/\sqrt{\om})$, in agreement with~\cite{Macher09b}. This means that conversion from short to long wavelength mode, and in particular spontaneous emission (governed by $\beta$), becomes very large at small frequencies, while conversion between long wavelength modes is bounded. We underline that this holds for \emph{arbitrary} backgrounds. The only necessary point is the asymptotic behavior: the flow must make a transition from subsonic to supersonic.

The second mode we will build is the negative energy long wavelength mode coming out to the left, i.e. $\theta_\hp^{\rm out}$. In the region $L \ll |x| \ll c/\om$, this mode reads
\bsub \label{thetaHP} \bea
\theta_\hp^{\rm out} &\underset{-\infty}{\sim}& \sqrt{\frac{m c_l}{2\om \rho_l}} - i\frac{\sqrt{m c_l \om}}{(v_l - c_l)\sqrt{2 \rho_l}}x + \tilde \alpha^* U_Z e^{i q_l x} - \tilde \beta^* U_Z e^{-i q_l x} , \label{thetaHP_left} \\
&\underset{+\infty}{\sim}& -B^* \sqrt{\frac{m c_r}{2\om \rho_r}} + i\frac{B^* \sqrt{m c_r \om}}{(c_r + v_r)\sqrt{2\rho_r}}x .
\eea \esub
To identify this mode to a low-frequency solution of \eq{mode_eq}, we need to build a solution that is linearly independent from $\theta_1$. Then we will identify \eqref{thetaHP} to the linear combination
\be \label{CombiLi}
\mathcal N^1 \theta_1 + \mathcal N^2 \theta_2
\ee
To do so, we first notice that for $\om = 0$, the couple $n = 0$ and $\theta = $const. is a solution of \eq{mode_eq}. Since $\theta_1$ has no constant term on the left, we see that in order to have the same asymptotic as in \eqref{thetaHP_left}, we can choose $\theta_2 = \sqrt{m c_l/(2\om \rho_l)}$, which fixes $\mathcal N^2$ to 1. This is, however, not enough. Indeed, as previously, we must obtain the first-order correction in $\om$. To do so, we write $\theta_2$ under the form
\be
\theta_2 = \sqrt{\frac{m c_l}{2\om \rho_l}} \left( 1 + i \om \eps(x) \right).
\ee
Now, by a calculation very similar to the one to obtain \eq{First_om_corr}, we show that
\be
\frac{v^2 - c^2}{v} \p_x \eps = 1 + a c^2,
\ee
where $a$ is a constant of integration. We now fix the value of $a$ by taking $x \to -\infty$, and identifying $\p_x \eps$ to the $O(x)$ coefficient in \eq{thetaHP_left}. Taking the limit $x \to +\infty$ with that value of $a$ gives us the asymptotic behavior of $\theta_2$. This leads to
\bsub \bea
\theta_2 &\underset{-\infty}{\sim}& \sqrt{\frac{m c_l}{2\om \rho_l}}\left(1 - i\frac{\om x}{(v_l - c_l)} \right) , \\
&\underset{+\infty}{\sim}& \sqrt{\frac{m c_l}{2\om \rho_l}}\left(1 - i \om A_5 x \right),
\eea \esub
with
\be
A_5 = \frac{v_r}{v_r^2 - c_r^2} \left(1 + \frac{c_r^2}{v_l c_l}\right).
\ee
What is left to do is now to identify the linear combination \eqref{CombiLi} to the low-frequency mode \eqref{thetaHP}. This gives us again a $4\times 4$ linear system, which we invert to obtain
\bsub \bea
B &=& \frac{v_\eff^2 - v_l c_l}{v_\eff^2 + v_r c_r} \sqrt{\frac{v_r c_r}{v_l c_l}}, \label{B_eq} \\
\tilde \alpha &=& i \frac{A_2}{q_l \rho_l U_Z A_1} \sqrt{\frac{m c_l}{2 \om \rho_l}} \left(1 + B \sqrt{\frac{v_r c_r}{v_l c_l}}\right) , \\
\tilde \beta &=& i \frac{A_3}{q_l \rho_l U_Z A_1} \sqrt{\frac{m c_l}{2 \om \rho_l}} \left(1 + B \sqrt{\frac{v_r c_r}{v_l c_l}}\right).
\eea \esub
To obtain the last mode, we need to identify a different linear combination \eqref{CombiLi} to
\bsub \bea
\theta_\v^{\rm out} &\underset{-\infty}{\sim}& \sqrt{\frac{m c_l}{2\om \rho_l}} - i\frac{\sqrt{m c_l \om}}{(v_l + c_l)\sqrt{2\rho_l}}x + \tilde R^* U_Z e^{i q_l x} - \tilde B^* U_Z e^{-i q_l x} , \\
&\underset{+\infty}{\sim}& \tilde T \sqrt{\frac{m c_r}{2\om \rho_r}} - i\frac{\tilde T \sqrt{m c_r \om}}{(c_r + v_r)\sqrt{2\rho_r}}x.
\eea \esub
We can proceed in exactly the same way as above. However, it is quicker to notice that $\theta_\v^{\rm out}$ can be obtain from $\theta_\hp^{\rm out}$ by the formal replacement $c_l \to -c_l$. Either way, we obtain the rest of the scattering coefficients
\bsub \bea
\tilde T &=& \frac{v_\eff^2 + v_l c_l}{v_\eff^2 + v_r c_r} \sqrt{\frac{v_r c_r}{v_l c_l}}, \\
\tilde R &=& i \frac{A_2}{q_l \rho_l U_Z A_1} \sqrt{\frac{m c_l}{2 \om \rho_l}} \left(1 - \tilde T \sqrt{\frac{v_r c_r}{v_l c_l}}\right) , \label{Rtilde} \\
\tilde B &=& i \frac{A_3}{q_l \rho_l U_Z A_1} \sqrt{\frac{m c_l}{2 \om \rho_l}} \left(1 - \tilde T \sqrt{\frac{v_r c_r}{v_l c_l}}\right). \label{Btilde}
\eea \esub
We see here that the coefficients $\tilde R$ and $\tilde B$, although they also describe coupling to the mode $\theta_\v$, behaves quite differently from $R$ and $B$, and scale like $O(1/\sqrt{\om})$. The reason is that $R$ and $B$ describes the coupling between the long wavelength modes, while $\tilde R$ and $\tilde B$ encode the production of the long wavelength mode $\theta_\v$ when sending a dispersive mode $\theta_\pm$. In addition, although the expressions of $\tilde R$ and $\tilde B$ are quite similar to $\tilde \alpha$ and $\tilde \beta$, they differ in an important respect. When $\theta_\v$ decouples, $R$ and $B$ becomes small, while $\tilde T$ becomes close to 1. In fact, the combination $1 - \tilde T \sqrt{v_r c_r/v_l c_l}$ becomes small. This means that when the coupling to $\theta_\v$ is small, $\tilde R$ and $\tilde B$ are significantly smaller that $\tilde \alpha$ and $\tilde \beta$. This is possible because $\tilde T$ also scales as $O(\om^0)$, as $R$ and $B$, since it also encodes a transition from a long wavelength mode to itself. As a last remark, we point out that the three coefficients $R$, $B$ and $\tilde T$ are related by norm conservation. Indeed, normalization of the third line of $S$ imposes
\be
|\tilde T|^2 - |B|^2 + |R|^2 = 1,
\ee
which can be checked explicitly to hold.

\section{Scattering coefficients}
\label{Coefs_Sec}
\subsection{Behavior of greybody factors}
There are two possibilities to minimize the effects of the greybody factors $R$ and $B$. The first one is to approach the \emph{near-critical} regime. When the flow speed is close to the sound speed on both sides, i.e. $v_r \sim c_r \sim v_l \sim c_l$, the mode $\theta_\v$ \emph{universally} decouples. More precisely, all greybody factors decrease as $O(c-v)$. The second possibility is to tune the variations of $v$ and $c$ (which can be achieved by controlling $V_{\rm ext}(x)$ and $g(x)$ as mentioned in section~\ref{Bckgr_Sec}) so that $R$ and $B$ vanish.

We saw in the preceding section that using the effective velocity $v_\eff$, the greybody factors have a simple and universal form. In particular $R$ and $B$ only depend on dispersion and the details of the profile such as its length $L$, through $v_\eff$. It is therefore useful to start by characterizing this effective velocity. The first point to notice is that in the smooth limit $L \gg \xi$, $v_\eff$ becomes equal to the absolute value of the velocity at the horizon $v_0 \equiv -v(x_{\mathcal H})$, i.e.
\be \label{veff_rel}
v_\eff \underset{L \gg \xi}\sim v_0,
\ee
This is because, in the limit $L \gg \xi$, the mode $\theta_\om$ is accurately given by WKB waves, except near the horizon where $v^2(x) = c^2(x)$, where there is a turning point. Hence the integral defining $v_\eff$ is mainly governed by the vicinity of that point (see appendix~\ref{veff_App} for more details). In the step limit $L \ll \xi$, $v_\eff$ can also be computed explicitly (this is done in appendix~\ref{veff_App}, see \eq{veff_step_App}), and we see that it is between $v_r$ and $v_l$, depending on how symmetric the flow is.

\begin{figure}[!ht]
\begin{center}
\includegraphics[width=0.49\columnwidth]{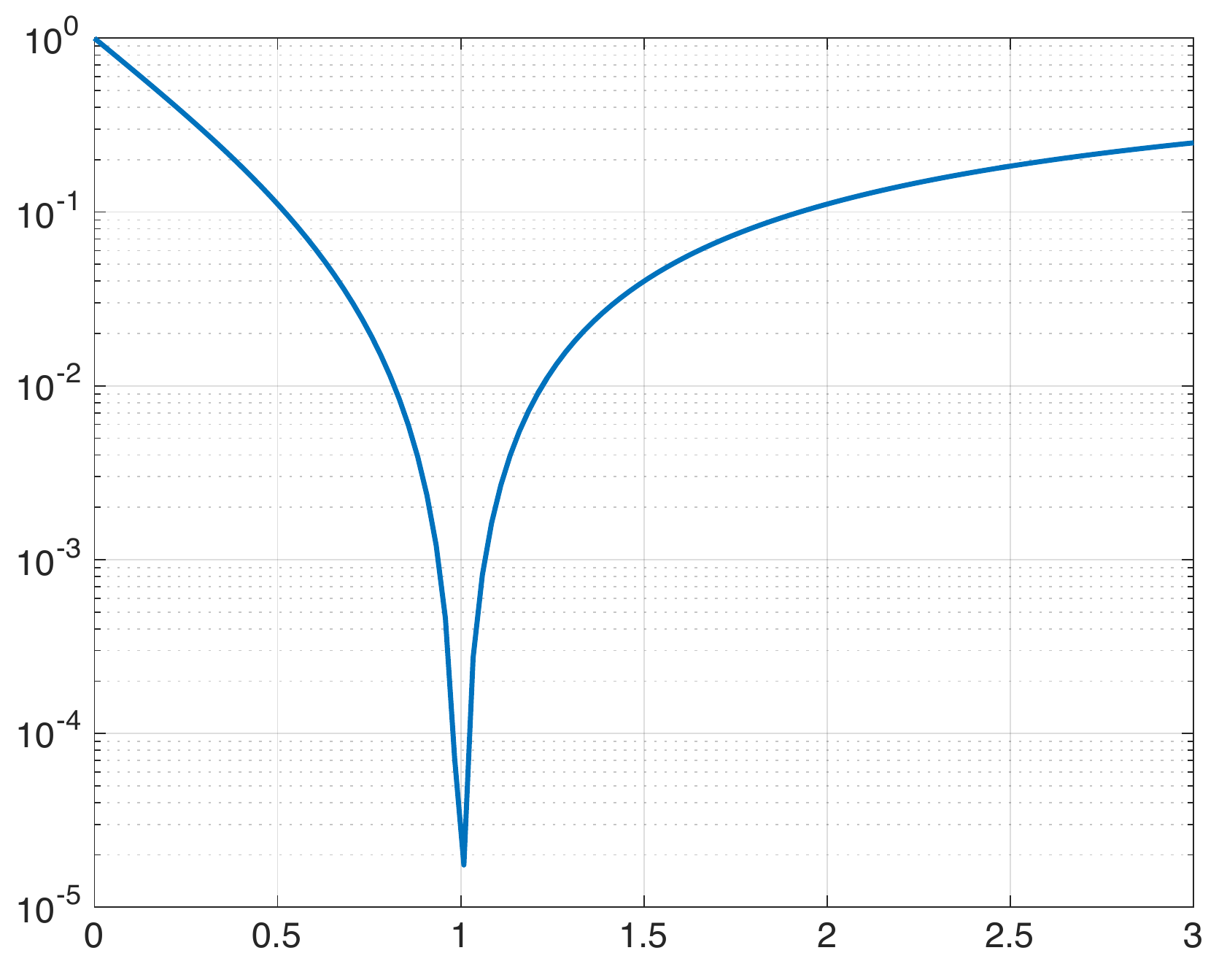}
\includegraphics[width=0.49\columnwidth]{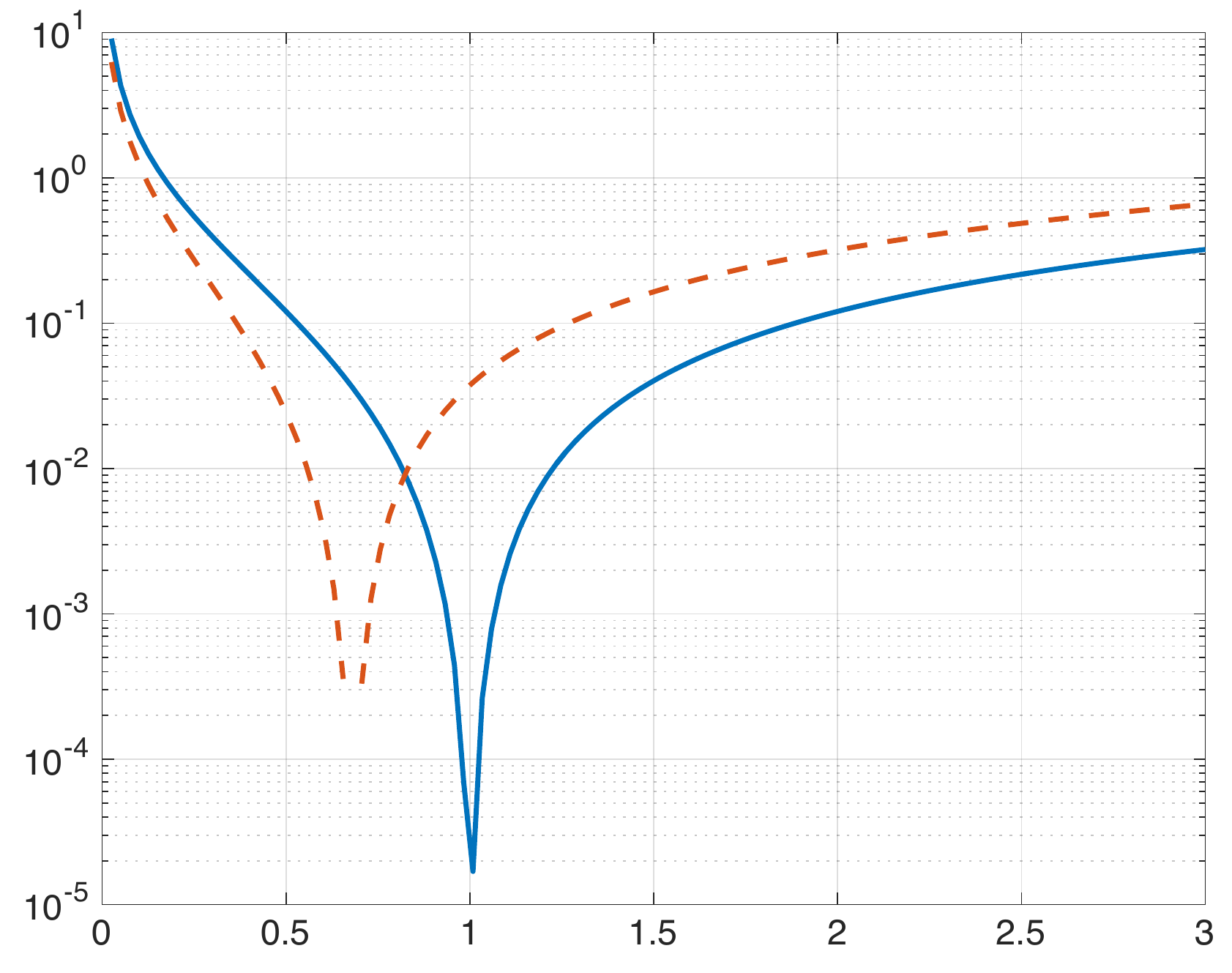}
\end{center}
\caption{Left panel: $\ln (|R|^2)$ as a function of $p_r$. Right panel: $\ln (|B|^2)$ (solid line) and $\ln |\zeta_{\u \u} |$ (dashed line) as a function of $p_l$, and for $p_r = 0.7$ (for which $R \neq 0$). The parameter $\zeta_{\u \u}$ is defined in \eq{CS_def}.
}
\label{R_Fig}
\end{figure}

The first important result from the expressions \eqref{R_eq} for $R$, and \eqref{B_eq} for $B$, is that the spectator mode $\theta_\v$ always decouple in the near-critical regime. Indeed, we see that if $v_r \sim c_r \sim v_l \sim c_l$, then $v_\eff \sim v_r$, and hence $R$ and $B$ vanish. In fact, in this regime, the scattering is accurately described by the \emph{Korteweg-de Vries} model, that we studied in~\cite{Coutant17}. This will be further confirmed in section~\ref{solvable_Sec}, where we show (in explicit examples) that the expressions for the other coefficients reduce to the one of the Korteweg-de Vries model. It is also possible to reduce significantly the mixing with $\theta_\v$ by carefully choosing the relative variations of $v$ and $c$, even outside the near-critical regime. Indeed, we see that $R = 0$ (resp. $B=0$) if we have $v_\eff^2 = v_r c_r$ (resp. $v_\eff^2 = v_l c_l$). To see this in more details, we first consider the smooth regime $L \gg \xi$. We introduce a new set of (mutually independent) parameters
\be \label{Mach_def}
M_{l/r} = \frac{v_{l/r}}{c_{l/r}}, \qquad \textrm{and} \qquad p_{l/r} = \frac{v_{l/r} c_{l/r}}{v_0^2}.
\ee
$M_{l/r}$ is the Mach number on the left or right side, while $p_{l/r}$ encodes how symmetrically $v$ and $c$ varies off their horizon value $v_0$. When $p=1$, inhomogeneities are symmetrically shared between $v$ and $c$. This parametrization is quite convenient, since $R$ and $B$ only depend on $p_{r/l}$, and not on the Mach numbers themselves. Using equations \eqref{R_eq}, and \eqref{B_eq}, we obtain
\bsub \label{RB_peq} \bea
R &=& \frac{1 - p_r}{1+p_r} , \\
B &=& \frac{1 - p_l}{1+p_r} \sqrt{\frac{p_r}{p_l}} .
\eea \esub
We first see from these equations that when $p_r = p_l = 1$, the two coefficients vanish, which means that $\theta_\v$ completely decouples (see Fig.~\ref{R_Fig}). Since our treatment only provides the coefficients up to higher powers of $\omega$, when these expressions vanish, we can only conclude that $R = O(\om L/c)$ (or $B$). When relaxing the assumption $L \gg \xi$, the coefficients $R$ and $B$ are still given by \eq{RB_peq}, but with $p_{r/l}$ defined with $v_\eff$ instead of $v_0$ in \eqref{Mach_def}. Hence, one can still obtain a vanishing $R$ (resp. $B$) if $v_\eff = v_r c_r$ (resp. $v_l c_l$).

In the smooth limit $L \gg \xi$, the fact that $\theta_\v$ decouples for $p_r = p_l = 1$ comes from a hidden symmetry of the phononic mode \eq{mode_eq}. It is known from early works in analogue gravity~\cite{Unruh81,Garay00} that the phononic wave equation of a three-dimensional fluid reduces to the wave equation in a Lorentzian geometry. This is, however, not the case in a one-dimensional fluid~\cite{Barcelo05}. A key difference is that the wave equation in a 1+1 dimensional space-time is conformally invariant while the phononic wave equation is not. However, it turns out that the two coincide under the condition that 
\be \label{conformal_cond}
v(x) c(x) = \mathrm{const}.
\ee
This implies that the phononic equation becomes \emph{conformally invariant} when this condition is met. In particular, $\theta_\v$ exactly decouples, and hence the greybody factors $R$ and $B$ become trivial for all $\om$. This property of the phononic wave equation is shown in details in appendix~\ref{conformal_App}.

\subsection{Effect of temperature and consequences on the observation of entanglement}
\label{Entang_Sec}
In realistic conditions, the condensate has a finite temperature. The incoming dispersive modes have typically a short wavelength, shorter that the thermal wavelength, and hence it is reasonable to assume that they are in their ground state (but see e.g.~\cite{Busch14} for more general states). On the other hand, the incoming long wavelength mode is in general thermally excited. Hence we assume that the incoming flux $n_\v^{\rm in}$ of this mode follows a Planck distribution at the temperature $T_\v$:
\be
n_\v^{\rm in} = \frac1{e^{\om/T_\v} - 1} \underset{\om \ll T_\v}\sim \frac{T_\v}{\om}.
\ee
In principle, $T_\v$ is given by the ambient temperature $T_{\rm ext}$ multiplied by a Doppler factor due to the flow, $T_\v = T_{\rm ext}(1+|v|/c)$. In this state (vacuum for $\theta_\pm^{\rm in}$ and thermal for $\theta_\v^{\rm in}$), the emitted fluxes of phonons are given by
\bsub \bea
n_\u &=& |\beta|^2 + n_\v^{\rm in} |R|^2 , \\
n_\hp &=& |\tilde \beta|^2 + (1 + n_\v^{\rm in}) |B|^2 , \\
n_\v &=& |\tilde B|^2 + n_\v^{\rm in} |\tilde T|^2 .
\eea \esub
Moreover, an initial temperature has also a significant effect on the correlation properties of the various modes. In the Hawking effect, the emitted mode $\theta_\u$ and its partner $\theta_\hp$ are quantum mechanically entangled~\footnote{This can be carefully defined using e.g. the notion of non-separable states~\cite{Werner89} for bipartite systems. A violation of the Cauchy-Schwarz inequality (\eq{CSineq}) is a sufficient condition to show that a state is non-separable.}. A convenient criterion to show this entanglement is to compare autocorrelations with cross-correlations~\cite{Busch14,Finke16,Michel16}. Correlations between classical excitations must obey the following bound, known as the Cauchy-Schwarz inequality:
\be \label{CSineq}
|c_{\u \u}|^2 \equiv \left|\mathrm{Tr}\left(\hat \rho \hat a_\u \hat a_\hp\right)\right|^2 < n_\u n_\hp,
\ee
where $\hat \rho$ is the density matrix of the phononic state. In~\cite{Steinhauer15}, this criterion was used to assess the entanglement in an experimental realization of the Hawking effet in an Bose-Einstein condensate. At zero temperature $T_{\rm ext} = 0$, this inequality is violated, showing that the pairs are entangled. This changes when one takes into account a temperature for the incoming mode $\theta_\v$~\cite{Macher09b,Busch14}. In that case, using \eq{aaS} the difference between auto-correlations and cross-correlations is given by~\footnote{We refer to~\cite{Busch14} for more details, see in particular equations (7) and (22).}
\be \label{CS_expr}
n_\u n_\hp - |c_{\u \u}|^2 = \left(|\tilde \beta|^2 - |\beta|^2\right) n_\v^{\rm in} - |\beta|^2.
\ee
To analyze the influence of the downstream mode on the Cauchy-Schwarz inequality \eqref{CSineq}, we introduce a parameter $\zeta_{\u \u}$ defined as
\be \label{CS_def}
\zeta_{\u \u} = \frac{|\tilde \beta|^2 - |\beta|^2}{|\beta|^2}.
\ee
We see from \eqref{CS_expr} that depending on the sign of $\zeta_{\u \u}$, the coupling with the mode $\theta_\v$ increases ($\zeta_{\u \u} > 0$), or decreases ($\zeta_{\u \u} < 0$) the possibility of observing entanglement. Using the general expressions we obtained in equations \eqref{thetaU} and \eqref{thetaHP}, we see that this parameter is essentially governed by the greybody factors, since it can be expressed as
\be \label{CSparameter}
\zeta_{\u \u} = \frac{\left(\sqrt{\frac{v_l c_l}{v_r c_r}} + B\right)^2 - \left(1 + R\right)^2}{\left(1 + R\right)^2}.
\ee
Using the notations of \eq{Mach_def}, this simplifies into
\be \label{CSp}
\zeta_{\u \u} = \frac{(p_r-p_l)^2}{4p_r p_l} = \frac{(v_r c_r - v_l c_l)^2}{4 v_r v_l c_r c_l}.
\ee
This expression has several consequences. First, $\zeta_{\u \u}$ is manifestly positive. This means that the coupling to the mode $\theta_\v$ always reduces the possibility of observing entanglement through the Cauchy-Schwarz inequality. Second, $\zeta_{\u \u}$ is independent of $v_\eff$, and in particular of the dispersive scale $\xi$. As a result, $\zeta_{\u \u}$ \emph{does} vanish if $R$ and $B$ do, e.g. in a smooth flow with the conformal condition \eqref{conformal_cond}, but it also vanishes under the simpler condition that
\be \label{CS_cond}
v_r c_r = v_l c_l.
\ee
In other words, one can minimize, and even cancel the effect of the downstream mode $\theta_\v$ on the Cauchy-Schwarz inequality even if the greybody factors are non-zero, if their contributions compensate in \eq{CSparameter}. This can be achieved by enforcing the asymptotic values of $v$ and $c$ to satisfy \eqref{CS_cond}. Since this cancelling of $\zeta_{\u \u}$ is shown for small frequencies, the condition \eqref{CS_cond} implies in general $\zeta_{\u \u} = O(\om L/c)$. Moreover, under the \emph{conformal condition} \eqref{conformal_cond}, more constraining than \eqref{CS_cond}, higher powers of $\om$ are suppressed by the dispersive scale due to conformal invariance (see appendix~\ref{conformal_App}), that is by a factor $O(\xi/L)$. This means that under this latter condition, $\zeta_{\u \u}$ is highly suppressed, scaling as $O(\om \xi/c)$. The variations of $\zeta_{\u \u}$ with $p_{r/l}$ are shown in Fig.~\ref{R_Fig}.


\subsection{Exactly solvable examples}
\label{solvable_Sec}
\subsubsection{Constant flow velocity}
As a first example, we consider the case where the density $\rho = \rho_0$ and velocity $v = -v_0$ are constant, and all the variation is contained in $c(x)$ (and hence in $g(x)$). Although this is presumably difficult to set experimentally, this case is interesting for its mathematical simplicity. For this reason, it has been previously studied in details in the phononic regime~\cite{Anderson14,Anderson15,Fabbri16} and in the step regime~\cite{Fabbri10,Mayoral10}. Our method allows us to generalize these results for any value of the transition size $L$ compared to the healing length $\xi$. As mentioned above, since $v$ is constant, it immediately follows that $v_\eff = v_0$. Hence, the greybody factors are fixed irrespectively of the profile $c(x)$. Explicitly, we find
\bsub \bea
R &=& \frac{v_0 - c_r}{v_0 + c_r} , \\
B &=& -\frac{v_0 - c_l}{v_0 + c_r} \sqrt{\frac{c_l}{c_r}} , \\
\zeta_{\u \u} &=& \frac{(c_r-c_l)^2}{4c_r c_l} .
\eea \esub
As we saw in the previous section, we observe here one of the disadvantage of having all inhomogeneities in $c$. Indeed, $R$, $B$, and $\zeta_{\u \u}$ vanish only in the near critical limit. As discussed above, fine-tuning $R$, $B$, or $\zeta_{\u \u}$ to be 0 requires to vary both $v$ and $c$ in a specific manner. We now turn to the computation of the other scattering coefficients. When $v$ is constant, the \eq{Schro_eq} for the auxiliary field $\conj$ reduces to
\be
4 m^2 (c^2 - v_0^2) \conj = \p_x^2 \conj.
\ee
We now consider the profile
\be \label{flow1}
c^2(x) = c_r^2 + \frac{c_l^2 - c_r^2}{1+e^{x/L}} .
\ee
With it, \eq{Schro_eq} is exactly solvable in terms of hypergeometric functions. As we saw in \Sec{Gen_results_Sec}, to obtain the scattering coefficients we need the decaying mode solution of \eq{Schro_eq}. It is given by
\be \label{chi_ex1}
\conj_1 = e^{-q_r x} {}_2 F_1\left(q_r L - i q_l L, q_r L + i q_l L; 1 + 2q_r L; -e^{-x/L}\right).
\ee
Using known identities of hypergeometric function (see appendix~\ref{Hyper_App} for details), we obtain the asymptotic coefficients $A_{1,2,3}$ defined in equations~\eqref{chi_gen}, and \eqref{A1}. We shall now focus on the explicit expression of $\beta$. As we saw in \eq{CSparameter}, $\tilde \beta$ (and hence $\alpha$, $\tilde \alpha$) are then readily obtained using the expressions for $R$ and $B$, which we obtained above. Using the results of appendix~\ref{Hyper_App} and \eq{beta_eq}, it follows
\be \label{beta1brut}
|\beta|^2 = \frac{(v_0^2 - c_l^2) q_r^2}{2 \pi \om L (q_r^2 + q_l^2)} \times \frac{c_r \sinh(\pi q_l L)^2}{v_0 \sinh(2\pi q_l L)} \times (1+R)^2.
\ee
To compare this with the Hawking result of equations \eqref{HR} and \eqref{THR}, we evaluate the surface gravity of our background. We first determine using \eqref{flow1} where $c^2 = v_0^2$. We then evaluate the gradient at this location to deduce
\be
2v_0 \kappa = \p_x(c^2)_{\mathcal H} = \frac{(v_0^2 - c_l^2) (c_r^2 - v_0^2)}{L (c_r^2 - c_l^2)}.
\ee
We use this to rearrange the expression for $\beta$ above. We finally obtain
\be \label{BetaEx1}
|\beta|^2 = \frac{T_H}{\om} \times \tanh(\pi q_l L) \times \frac{4 v_0 c_r}{(v_0 + c_r)^2}.
\ee
This formula is the product of 3 terms with a clear physical interpretation. The first term is the Hawking formula at low frequencies (see \eqref{HR0}). The second term encodes how dispersion alter the Hawking temperature. In the limit $L \gg \xi$, we have $\tanh(\pi q_l L) \sim 1$, and the result become independent of the dispersive scale $\xi$. The last term is the modification of the emitted flux due to the coupling to the mode $\theta_\v$. It is equal to $1-|R|^2$, and corresponds to the fraction of long wavelength modes that are transmitted from the horizon to infinity. In the smooth regime $L \gg \xi$, \eq{BetaEx1} is the product of the low-frequency Planck spectrum with the (relativistic) transmission coefficient from the horizon to infinity~\cite{Anderson14,Anderson15}. This is the general result for a black hole in general relativity~\cite{Page76}. To summarize this behavior, it is tempting to define an effective temperature $T_\eff$, such that
\be
|\beta|^2 = \frac{T_\eff}{\om} \left(1 - |R|^2\right).
\ee
In the smooth limit, $T_\eff \sim T_H$, and one recovers the factorization of the relativistic case. However, since the coefficient $\beta$ always grows like $1/\sqrt{\om}$ at low frequencies, this equation can be seen as a definition of an effective temperature $T_\eff$ that takes into account greybody factors. In the near critical limit, $T_\eff$ has the value found in the Korteweg-de Vries model, as expected.

\subsubsection{Varying flow velocity and speed of sound}
\begin{figure}[!ht]
\begin{center}
\includegraphics[width=0.49\columnwidth]{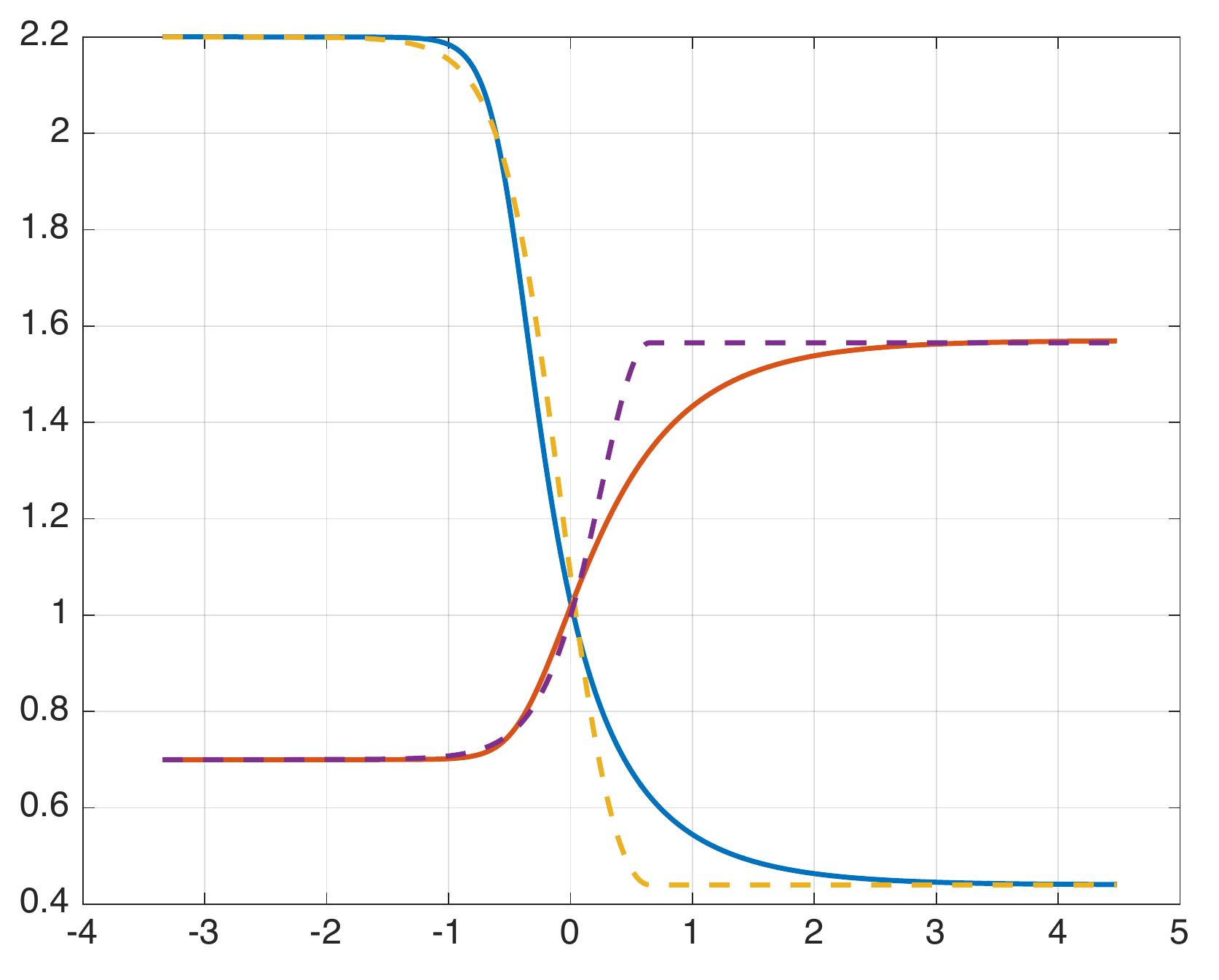}
\end{center}
\caption{Profile of $v$ (blue), and $c$ (red) as functions of $x/L$, obtained from equations \eqref{flow2} and \eqref{flow2b}. We have used the parameters $v_l = 2.2$, $c_l = 0.7$, $c_r = 1.57$, $v_r = 0.44$, and $\Delta/L = -5.7$. To compare with other works, we also plotted a waterfall solution (dashed curves), with $M_l = 5$~\cite{Larre12,Michel16}.
}
\label{WaterFall_Fig}
\end{figure}

To discuss the generality of the conclusions just drawn, it is interesting to look for another exactly solvable example where both $v$ and $c$ vary. This is slightly harder due to the presence of the operator $\p_x \rho \p_x$ in \eq{Schro_eq}. However, it is possible to construct a profile with similar solution as for \eqref{flow1} by tuning how $v$ and $c$ vary together, but still having their asymptotic values unspecified. For this, we first introduce a new spatial coordinate
\be
y = \int \frac{v(x)}{v_0} dx.
\ee
$y$ is normalized using $v_0$, the value of the velocity at the horizon, so it still has a dimension of length (this is a convenient choice, since gradients with respect to $x$ or $y$ are then equal on the horizon). Since $|v| \neq 0$ all along the flow, this change of variable is perfectly regular. Moreover, on both asymptotic sides, it simply amounts to a linear rescaling of $x$. Using this new coordinate, the equation \eqref{Schro_eq} for the field $\conj$ reads
\be \label{Schro_eq2}
4m^2 v_0^2 \left(\frac{c^2}{v^2} - 1\right) \conj = \p_y^2 \conj.
\ee
We see that what matters to obtain low-frequency solutions is the evolution of the Mach number with the new coordinate $y$. We assume the following profile
\be \label{flow2}
M^2(y) \equiv \frac{c^2(y)}{v^2(y)} = M_r^{-2} + \frac{M_l^{-2} - M_r^{-2}}{1+e^{y/L}}.
\ee
This profile is quite similar to \eqref{flow1}, but the asymptotic values of $v$ and $c$ are now independent. Notice that $v_0$ the value of the velocity at the horizon is also an independent parameter, since the profile of $v$ itself is not specified. To fix the ideas, we shall assume a similar shape as \eqref{flow2}, given by
\be \label{flow2b}
\frac1{v^2(y)} = \frac1{v_r^2} + \frac{v_l^{-2} - v_r^{-2}}{1+e^{(y - \Delta)/L}}.
\ee
The parameter $\Delta$ gives a shift between the center of the transition of the Mach number $M$ and that of $v$. Changing the value of $\Delta$ allows us to change the value of $v_0$. To illustrate what profiles are described by equations \eqref{flow2} and \eqref{flow2b}, we represented $v$ and $c$ in terms of the original spatial coordinate $x$ in Fig.~\ref{WaterFall_Fig}.
Similarly to the flow \eqref{flow1}, the solutions of \eq{Schro_eq2} are given by hypergeometric functions. In particular, the decaying mode simply reads
\be \label{chi_ex2}
\conj_1(y) = e^{-Q_r y} {}_2 F_1\left(Q_r L - i Q_l L, Q_r L + i Q_l L; 1 + 2Q_r L; -e^{-y/L}\right),
\ee
where we introduced the rescaled momenta
\bsub \bea
Q_r &=& \frac{q_r v_0}{v_r}, \\
Q_l &=& \frac{q_l v_0}{v_l} .
\eea \esub
Using these rescaled momenta, the expressions of $A_{1..4}$ are identical to the one in appendix~\ref{Hyper_App}. It follows a similar expression for $\beta$ than \eqref{beta1brut}. To again write it in a convenient way to compare with the Hawking result, we first evaluate the surface gravity of the flow, that is
\be
\frac{2 \kappa}{v_0} = \p_x((c/v)^2)_{\mathcal H} = \frac{(M_l^2 - 1)(1 - M_r^2)}{L(M_l^2 - M_r^2)}.
\ee
This leads to
\be \label{BetaEx2}
|\beta|^2 = \frac{T_H}{\om} \times \tanh(\pi Q_l L) \times \frac{4 v_r c_r v_\eff^4}{v_0^2 (v_\eff^2 + v_r c_r)^2}.
\ee
This formula has the same structure as \eqref{BetaEx1}, with 3 factors with the same interpretation. As previously, in the smooth limit $L \gg \xi$, $|\beta|^2$ reduces to the product of a Planck spectrum at temperature $T_H$ with the relativistic transmission coefficient from the horizon to infinity.

As a last remark, we point out that the second factor in \eqref{BetaEx2}, which governs the dispersive corrections to the Hawking temperature, becomes 1 when $\pi Q_l L \gg 1$. In particular, the dispersive corrections are sensitive to the properties of the flow on the \emph{supersonic} side. This is physically reasonable, since the dispersive modes live on that side~\cite{Coutant11}. In addition, $Q_l L = \sqrt{M_l^2 - 1} \times L/\xi_0$, and therefore, if the supersonic side has a large Mach number, the temperature will be close to the one predicted by Hawking, even if the condition $\xi/L \ll 1$ is not well satisfied. For instance, in the experiment of~\cite{Steinhauer15}, the length of the transition $L$ is essentially $\xi_0$, but because $M_l$ is larger than about 3, the temperature is expected to be quite close to the one predicted by Hawking, as was noticed in~\cite{Michel16}.

\section{Conclusion}

In this work we studied the analogue Hawking effect in Bose-Einstein condensates. As we show, the production of pairs of phonons \emph{via} the Hawking effect is dominant in the low-frequency regime. We developed a method to obtain analytical results for low frequencies in general flows. In a companion work~\cite{Coutant17}, we analyzed the dispersive corrections to the effective temperature of emitted phonons. In this work instead, we focused on the influence of the mode propagating with the flow. This mode is not the one responsible for the Hawking effect, and is for this reason sometimes called ``spectator,'' but it affects the observables through its coupling to the Hawking mode and its partner. In black hole physics, this mode gives rise to the so-called greybody factors. We studied here the generalization of these greybody factors in the Bogoliubov-de Gennes model.

Our first result is the general low-frequency dependence of all scattering coefficients. In particular, the spectrum of emitted phonons increases like the inverse power of the frequency (see \eq{beta_eq}), irrespectively of the coupling to the downstream mode. This is due to the fact that greybody factors tends to a constant in this regime, contrary to what happens in higher than $1+1$ dimensional black holes~\cite{Page76}. This conclusion was reached in~\cite{Anderson14,Anderson15,Fabbri16} using the wave equation in an acoustic metric, and is generalized here by taking into account dispersive effects. When the size of the transition from a subsonic to a supersonic flow is large compared to the healing length, the spectrum is  given by the curved space-time prediction (see equations \eqref{BetaEx1}, \eqref{BetaEx2}). More precisely, it is given by the product of the thermal spectrum with the greybody factors obtained from the relativistic equation.

Our method allows us to quantify the coupling to the downstream mode in general flows. In particular, we identified two distinct regimes where this coupling is small. First, it is always small for near-critical flows, that is, when $v \sim c$ on both side of the transition. In this regime, the Hawking effect is well described by a simpler model, namely the linear Korteweg-de Vries equation~\cite{Coutant17}. Second, when the product of the velocity flow with the speed of waves is constant (\eq{conformal_cond}), this coupling vanishes for the dispersionless equation. This means that in smooth flows, it will remain small, scaling like the ratio between the healing length and the size of the transition. This property is due to the conformal invariance of the dispersionless equation when the condition \eqref{conformal_cond} is met (this is detailed in appendix~\ref{conformal_App}). We also show that the parameter $\xi_{\u \u}$ controlling the influence of the downstream mode on entanglement vanishes under a weaker condition (see \eq{CS_cond}).

In experimental setups, if one can control with enough precision the external potential $V_{\rm ext}$ and the effective 1D coupling constant $g$, one can reduce the coupling to the downstream mode by ensuring that the flow velocity $v$ and the speed of sound $c$ are related in the appropriate way. Reducing this coupling has several advantages. First, the flux of emitted phonons becomes very close to the Planck law predicted by Hawking. Second, and perhaps more importantly, it ensures that the Hawking mode and its partner are entangled for a large range of frequencies. Indeed, the downstream mode tends to reduce their entanglement \emph{via} thermal effects or other sources of noise. Ensuring its decoupling significantly eases the detection of entanglement and the observation of violations of classical inequalities.

\acknowledgments 
We would like to thank Renaud Parentani for useful comments about the final version of this manuscript. This project has received funding from the European Union's Horizon 2020 research and innovation programme under the Marie Sk\l odowska-Curie grant agreement No 655524. S.~W.~acknowledges financial support provided under the Royal Society University Research Fellow (UF120112), the Nottingham Advanced Research Fellow (A2RHS2), the Royal Society Project (RG130377) grants and the EPSRC Project Grant (EP/P00637X/1).

\newpage
\appendix
\section{Asymptotic properties of hypergeometric functions}
\label{Hyper_App}
In this appendix, we recall several functional identities~\cite{Olver,DLMF} used in the core of the text. We defined hypergeometric functions as
\be
{}_2 F_1(a,b;c;z) = \sum_{n=0}^\infty \frac{(a)_n (b)_n}{(c)_n n!} z^n = \sum_{n=0}^\infty \frac{\Gamma(a+n) \Gamma(b+n) \Gamma(c)}{\Gamma(a) \Gamma(b) \Gamma(c+n) n!} z^n.
\ee
In the text, solutions of the second-order differential equation are given in terms of hypergeometric functions. To obtain their asymptotic behavior, we use the transformations of variables
\bea \label{TransVar}
{}_2 F_1(a,b;c;z) &=& \frac{\Gamma(c) \Gamma(b-a)}{\Gamma(b) \Gamma(c-a)} (-z)^{-a} {}_2 F_1(a,a-c+1;a-b+1; z^{-1}) , \nonumber \\
&&+ \frac{\Gamma(c) \Gamma(a-b)}{\Gamma(a) \Gamma(c-b)} (-z)^{-b} {}_2 F_1(b,b-c+1;b-a+1; z^{-1}).
\eea
Using it with the solution of \eq{chi_ex1}, we obtain
\bsub \bea
A_2 &=& \frac{\Gamma(1+2 q_r L) \Gamma(2 i q_l L)}{\Gamma(q_r L + i q_l L) \Gamma(1 + q_r L + i q_l L)} , \\
A_3 &=& \frac{\Gamma(1+2 q_r L) \Gamma(-2 i q_l L)}{\Gamma(q_r L - i q_l L) \Gamma(1 + q_r L - i q_l L)} .
\eea \esub
Similarly, we use the identity
\be
\int_0^{\infty} z^{d-1} {}_2 F_1(a,b;c;-z) dz = \frac{\Gamma(d) \Gamma(c) \Gamma(a-d) \Gamma(b-d)}{\Gamma(a) \Gamma(b) \Gamma(c-d)},
\ee
to obtain $A_1$ from its definition \eqref{A1}. Using the solution \eqref{chi_ex1} in the integral \eqref{A1}, we find
\be
A_1 = \frac{\pi \Gamma(1+ 2 q_r L)}{q_r q_l \rho_0 L \sinh\left(\pi q_l L\right) \left|\Gamma(q_r L + i q_l L)\right|^2}.
\ee
From this we obtain the ratios that are relevant for the various scattering coefficients, namely,
\be
\frac{A_2}{A_1} = \frac{A_3^*}{A_1} = \frac{q_r q_l \rho_0 \Gamma(2 i q_l L) \sinh(\pi q_l L)}{\pi (q_r + i q_l)} \times \frac{\left|\Gamma(q_r L + i q_l L)\right|^2}{\Gamma(q_r L + i q_l L)^2} .
\ee

\section{Characterization of the effective velocity $v_\eff$}
\label{veff_App}

In the smooth limit $L \gg \xi$, the solutions of the auxiliary field equation \eqref{Schro_eq} are well approximated by WKB waves
\be
\chi(x) \sim \frac{e^{i \int k(x') dx'}}{\sqrt{|k(x)/v(x)|}},
\ee
where $k^2(x) = 4m^2 (c^2 - v^2)$ (and the extra $v$ factor in the amplitude comes from having $\rho^{-1} \p_x \rho \p_x$ instead of the standard $\p_x^2$). At the horizon, at the location $x_{\mathcal H}$, we have a turning point, and the WKB wave is singular. This means that on the left side, it is a superposition of oscillatory waves, while it decay exponentially on the other side, similarly to \eq{chi_gen}. In the WKB limit, the amplitudes on each side can be obtained using a connection formula~\cite{Berry72}. Fortunately, this is not necessary here. Indeed, for a simple turning point (which is the case considered here) the singularity of the WKB wave is integrable, since $|\chi| = O((x-x_{\mathcal H})^{-1/2})$. This is enough to evaluate both $A_1$ and $A_4$, both given by integrals of $\chi$ (or $\p_x \theta_\om$, see \eqref{A1} and \eqref{A4}). We must then evaluate an integral with a slowly varying amplitude and a rapidly varying phase. In this regime, the integral is dominated by its boundary, i.e. the turning point $x_{\mathcal H}$. Therefore
\be
\int_{-\infty}^{+\infty} \frac{\p_x \theta_{\om = 0}}{v^2} dx \sim \int_{x_{\mathcal H}} \frac{e^{i \int k(x') dx'}}{v^2 \rho \sqrt{|k/v|}} dx \sim \frac1{v_0^2} \int_{x_{\mathcal H}} \frac{e^{i \int k(x') dx'}}{\rho \sqrt{|k/v|}} dx \sim \frac1{v_0^2} A_1.
\ee
Using the definition \eqref{veff_def} of $v_\eff$, we deduce that in the smooth limit, $v_\eff \sim v_0$, i.e. \eq{veff_rel}. In the other limit, $L \ll \xi$, we can approximate the background with step function, e.g. $v(x) = -v_l \Theta(-x) - v_r \Theta(x)$. The \eq{Schro_eq} for $\chi$ can be directly solved. We see that
\be
\chi = \left\{\bal
& A_2 e^{i q_l x} + A_3 e^{-i q_l x} &\qquad (x<0), \\
& e^{- q_r x} &\qquad (x<0) .
\eal \right.
\ee
The amplitudes can be determined by imposing that both $\chi$ and $\rho \p_x \chi$ are continuous across $x = 0$. It is, however, enough to use the continuity of $\rho \p_x \chi$ to obtain
\be
A_2 - A_3 = i \frac{q_r \rho_r}{q_l \rho_l}.
\ee
We then compute $A_1$ from its definition \eqref{A1},
\bsub \bea
A_1 &=& \frac1{\rho_l} \int_{-\infty}^0 \left(A_2 e^{i q_l x} + A_3 e^{-i q_l x} \right) dx + \frac1{\rho_r} \int_0^{+\infty} e^{-i q_r x} dx , \\
&=& \frac1{i q_l \rho_l} (A_2 - A_3) + \frac1{q_r \rho_r}, \\
&=& \frac{q_l^2 \rho_l^2 + q_r^2 \rho_r^2}{q_l^2 \rho_l^2 q_r \rho_r},
\eea \esub
where we have used the prescription of footnote~\ref{regul_ftn}. Similarly for the integral defining $v_\eff$ in \eqref{veff_def}, we have
\be
\frac{A_1}{v_\eff^2} = \frac{v_l^2 q_l^2 \rho_l^2 + v_r^2 q_r^2 \rho_r^2}{v_l^2 v_r^2 q_l^2 \rho_l^2 q_r \rho_r}.
\ee
From this we obtain
\bsub \label{veff_step_App} \bea
v_\eff^2 &=& v_l^2 \frac{q_l^2}{q_r^2 + q_l^2} + v_r^2 \frac{q_r^2}{q_r^2 + q_l^2} , \\
&=& v_l^2 \frac{v_l^2 - c_l^2}{v_l^2 - c_l^2 + c_r^2 - v_r^2} + v_r^2 \frac{c_r^2 - v_r^2}{v_l^2 - c_l^2 + c_r^2 - v_r^2}.
\eea \esub
In particular, we see that $v_r < v_\eff < v_l$.

\section{Hidden symmetry of the phononic wave equation}
\label{conformal_App}

Starting from \eq{mode_eq}, it is easy to show that time dependent phase fluctuations $\theta(t,x) = \int a_\om \theta_\om e^{- i \om t} d\om$ in the dispersionless regime (i.e. dropping fourth-order derivatives) obey the equation
\be \label{CST_eq}
(\p_t + \p_x v) \frac{\rho}{c^2} (\p_t + v\p_x) \theta - \p_x \rho \p_x \theta = 0.
\ee
This equation can be identified as the one-dimensional reduction of a three-dimensional wave equation in a curved space-time. Indeed, starting from the metric described by the line element
\be \label{ds2}
ds^2 = \frac{\rho}{c} \left[c^2 dt^2 - (dx - vdt)^2 - dy^2 - dz^2\right] ,
\ee
the wave equation ($\p_\mu \sqrt{|g|} g^{\mu \nu} \p_\nu \theta = 0$ in relativistic notations) for fields independent of $y$ and $z$ reduces to \eqref{CST_eq}. This was shown initially for sound waves in a classical fluid~\cite{Unruh81}, and later for sound in a Bose gas in the Bogoliubov approximation~\cite{Garay00}. However, it is also known that \eq{CST_eq} does \emph{not} correspond to the one-dimensional wave equation in the metric
\be \label{ds3}
ds^2 = \frac{\rho}{c} \left[c^2 dt^2 - (dx - vdt)^2 \right] ,
\ee
Indeed, the 1+1 wave equation would be, instead of \eqref{CST_eq}, given by
\be \label{Conf_eq}
(\p_t + \p_x v) \frac{1}{c} (\p_t + v\p_x) \theta - \p_x c \p_x \theta = 0.
\ee
A key difference between this equation and \eqref{CST_eq}, is that \eq{Conf_eq} is \emph{conformally invariant}. This is a known property of the wave equation in 1+1 dimensions, and it can be seen in a relatively straightforward way. Indeed, \eq{Conf_eq} is equivalent to the set of (uncoupled) first-order equations
\bsub \bea
(\p_t + v\p_x) \theta_\u &=& c \p_x \theta_\u, \label{Conf_u} \\
(\p_t + v\p_x) \theta_\v &=& - c \p_x \theta_\v, \label{Conf_v}
\eea \esub
in the sense that every solution of \eqref{Conf_eq} is a sum $\theta = \theta_\u + \theta_\v$, where $\theta_\u$ obeys \eqref{Conf_u}, and $\theta_\v$ obeys \eqref{Conf_v}. This means that the general solution is a superposition of waves traveling upstream and waves traveling downstream, that don't interact with each other. In particular, the $S$-matrix is trivial, i.e. greybody factors exactly vanish. Note however that the Hawking effect is still present, due to the conformal anomaly~\cite{Christensen77}. Comparing the two equations, \eqref{CST_eq} and \eqref{Conf_eq}, we see that they coincide if and only if $\rho/c$ is a constant. Since in a Bose-Einstein condensate, the continuity equation imposes $\rho v = $ const, we conclude that the dispersionless equation for sound is conformally invariant under the condition
\be
v c = \mathrm{const}.
\ee
as was claimed in the core of the text.

\bibliographystyle{utphys}
\bibliography{Bibli}

\providecommand{\href}[2]{#2}\begingroup\raggedright\begin{thebibliography}{10}

\bibitem{Unruh81}
W.~Unruh, ``{Experimental black hole evaporation},''
\href{http://dx.doi.org/10.1103/PhysRevLett.46.1351}{{\em Phys. Rev. Lett.}
  {\bfseries 46} (1981) 1351--1353}.

\bibitem{Hawking75}
S.~Hawking, ``{Particle Creation by Black Holes},''
\href{http://dx.doi.org/10.1007/BF02345020, 10.1007/BF02345020}{{\em Commun.
  Math. Phys.} {\bfseries 43} (1975) 199--220}.

\bibitem{Weinfurtner10}
S.~Weinfurtner, E.~W. Tedford, M.~C. Penrice, W.~G. Unruh, and G.~A. Lawrence,
  ``{Measurement of stimulated Hawking emission in an analogue system},''
  \href{http://dx.doi.org/10.1103/PhysRevLett.106.021302}{{\em Phys. Rev.
  Lett.} {\bfseries 106} (2011) 021302},
\href{http://arxiv.org/abs/1008.1911}{{\ttfamily arXiv:1008.1911 [gr-qc]}}.

\bibitem{Belgiorno10}
F.~Belgiorno, S.~Cacciatori, M.~Clerici, V.~Gorini, G.~Ortenzi, L.~Rizzi,
  E.~Rubino, V.~Sala, and D.~Faccio, ``Hawking radiation from ultrashort laser
  pulse filaments,''
  \href{http://dx.doi.org/10.1103/PhysRevLett.105.203901}{{\em Phys. Rev.
  Lett.} {\bfseries 105} no.~20, (2010) 203901},
  \href{http://arxiv.org/abs/1009.4634}{{\ttfamily arXiv:1009.4634 [gr-qc]}}.

\bibitem{Euve15}
L.~P. Euv{\'e}, F.~Michel, R.~Parentani, T.~G. Philbin, and G.~Rousseaux,
  ``{Observation of noise correlated by the Hawking effect in a water tank},''
  \href{http://dx.doi.org/10.1103/PhysRevLett.117.121301}{{\em Phys. Rev.
  Lett.} {\bfseries 117} no.~12, (2016) 121301},
\href{http://arxiv.org/abs/1511.08145}{{\ttfamily arXiv:1511.08145
  [physics.flu-dyn]}}.

\bibitem{Steinhauer15}
J.~Steinhauer, ``{Observation of thermal Hawking radiation and its entanglement
  in an analogue black hole},'' \href{http://dx.doi.org/10.1038/nphys3863}{{\em
  Nature Phys.} {\bfseries 12} (2016) 959},
\href{http://arxiv.org/abs/1510.00621}{{\ttfamily arXiv:1510.00621 [gr-qc]}}.

\bibitem{Torres16}
T.~Torres, S.~Patrick, A.~Coutant, M.~Richartz, E.~W. Tedford, and
  S.~Weinfurtner, ``{Observation of superradiance in a vortex flow},''
\href{http://arxiv.org/abs/1612.06180}{{\ttfamily arXiv:1612.06180 [gr-qc]}}.

\bibitem{Jacobson91}
T.~Jacobson, ``{Black hole evaporation and ultrashort distances},''
\href{http://dx.doi.org/10.1103/PhysRevD.44.1731}{{\em Phys. Rev.} {\bfseries D
  44} (1991) 1731--1739}.

\bibitem{Macher09b}
J.~Macher and R.~Parentani, ``{Black hole radiation in Bose-Einstein
  condensates},'' \href{http://dx.doi.org/10.1103/PhysRevA.80.043601}{{\em
  Phys. Rev.} {\bfseries A 80} (2009) 043601},
\href{http://arxiv.org/abs/0905.3634}{{\ttfamily arXiv:0905.3634
  [cond-mat.quant-gas]}}.

\bibitem{Page76}
D.~N. Page, ``{Particle Emission Rates from a Black Hole: Massless Particles
  from an Uncharged, Nonrotating Hole},''
\href{http://dx.doi.org/10.1103/PhysRevD.13.198}{{\em Phys. Rev.} {\bfseries D
  13} (1976) 198--206}.

\bibitem{Mayoral10}
C.~Mayoral, A.~Fabbri, and M.~Rinaldi, ``{Step-like discontinuities in
  Bose-Einstein condensates and Hawking radiation: dispersion effects},''
  \href{http://dx.doi.org/10.1103/PhysRevD.83.124047}{{\em Phys. Rev.}
  {\bfseries D 83} (2011) 124047},
\href{http://arxiv.org/abs/1008.2125}{{\ttfamily arXiv:1008.2125 [gr-qc]}}.

\bibitem{Coutant13}
A.~Coutant and R.~Parentani, ``{Undulations from amplified low frequency
  surface waves},'' \href{http://dx.doi.org/10.1063/1.4872025}{{\em Phys.
  Fluids} {\bfseries 26} (2014) 044106},
\href{http://arxiv.org/abs/1211.2001}{{\ttfamily arXiv:1211.2001
  [physics.flu-dyn]}}.

\bibitem{Anderson14}
P.~R. Anderson, R.~Balbinot, A.~Fabbri, and R.~Parentani, ``{Gray-body factor
  and infrared divergences in 1D BEC acoustic black holes},''
  \href{http://dx.doi.org/10.1103/PhysRevD.90.104044}{{\em Phys. Rev.}
  {\bfseries D 90} no.~10, (2014) 104044},
\href{http://arxiv.org/abs/1404.3224}{{\ttfamily arXiv:1404.3224 [gr-qc]}}.

\bibitem{Anderson15}
P.~R. Anderson, A.~Fabbri, and R.~Balbinot, ``{Low frequency gray-body factors
  and infrared divergences: rigorous results},''
  \href{http://dx.doi.org/10.1103/PhysRevD.91.064061}{{\em Phys. Rev.}
  {\bfseries D 91} no.~6, (2015) 064061},
\href{http://arxiv.org/abs/1501.01953}{{\ttfamily arXiv:1501.01953 [gr-qc]}}.

\bibitem{Fabbri16}
A.~Fabbri, R.~Balbinot, and P.~R. Anderson, ``{Scattering coefficients and
  gray-body factor for 1D BEC acoustic black holes: exact results},''
  \href{http://dx.doi.org/10.1103/PhysRevD.93.064046}{{\em Phys. Rev.}
  {\bfseries D 93} no.~6, (2016) 064046},
\href{http://arxiv.org/abs/1512.08447}{{\ttfamily arXiv:1512.08447 [gr-qc]}}.

\bibitem{LandauV3}
L.~D. Landau and E.~M. Lifshitz, {\em Quantum mechanics: non-relativistic
  theory}, vol.~3, p.~81 and 245.
\newblock Elsevier, 2013.

\bibitem{Brito15}
R.~Brito, V.~Cardoso, and P.~Pani, ``{Superradiance},''
  \href{http://dx.doi.org/10.1007/978-3-319-19000-6}{{\em Lect. Notes Phys.}
  {\bfseries 906} (2015) pp.1--237},
\href{http://arxiv.org/abs/1501.06570}{{\ttfamily arXiv:1501.06570 [gr-qc]}}.

\bibitem{Coutant17}
A.~Coutant and S.~Weinfurtner, ``{Low-frequency analogue Hawking radiation: The
  Korteweg--de Vries model},''
  \href{http://dx.doi.org/10.1103/PhysRevD.97.025005}{{\em Phys. Rev.}
  {\bfseries D 97} no.~2, (2018) 025005},
\href{http://arxiv.org/abs/1707.09651}{{\ttfamily arXiv:1707.09651 [gr-qc]}}.

\bibitem{Garay00}
L.~Garay, J.~Anglin, J.~Cirac, and P.~Zoller, ``{Sonic black holes in dilute
  Bose-Einstein condensates},''
  \href{http://dx.doi.org/10.1103/PhysRevA.63.023611}{{\em Phys. Rev.}
  {\bfseries A 63} (2001) 023611},
\href{http://arxiv.org/abs/gr-qc/0005131}{{\ttfamily arXiv:gr-qc/0005131
  [gr-qc]}}.

\bibitem{Balbinot08}
R.~Balbinot, A.~Fabbri, S.~Fagnocchi, A.~Recati, and I.~Carusotto, ``{Non-local
  density correlations as signal of Hawking radiation in BEC acoustic black
  holes},'' \href{http://dx.doi.org/10.1103/PhysRevA.78.021603}{{\em Phys.
  Rev.} {\bfseries A 78} (2008) 021603},
\href{http://arxiv.org/abs/0711.4520}{{\ttfamily arXiv:0711.4520
  [cond-mat.other]}}.

\bibitem{Larre12}
P.~E. Larre, A.~Recati, I.~Carusotto, and N.~Pavloff, ``{Quantum fluctuations
  around black hole horizons in Bose-Einstein condensates},''
  \href{http://dx.doi.org/10.1103/PhysRevA.85.013621}{{\em Phys. Rev.}
  {\bfseries A 85} (2012) 013621},
\href{http://arxiv.org/abs/1110.4464}{{\ttfamily arXiv:1110.4464
  [cond-mat.quant-gas]}}.

\bibitem{Petrov00}
D.~Petrov, G.~Shlyapnikov, and J.~Walraven, ``Regimes of quantum degeneracy in
  trapped 1d gases,'' {\em Phys. Rev. Lett.} {\bfseries 85} no.~18, (2000)
  3745.

\bibitem{Pitaevskii}
L.~Pitaevskii and S.~Stringari, {\em Bose-Einstein condensation and
  superfluidity}, vol.~164.
\newblock Oxford University Press, 2016.

\bibitem{Coutant12}
A.~Coutant, A.~Fabbri, R.~Parentani, R.~Balbinot, and P.~Anderson, ``{Hawking
  radiation of massive modes and undulations},''
  \href{http://dx.doi.org/10.1103/PhysRevD.86.064022}{{\em Phys. Rev.}
  {\bfseries D 86} (2012) 064022},
\href{http://arxiv.org/abs/1206.2658}{{\ttfamily arXiv:1206.2658 [gr-qc]}}.

\bibitem{Dalfovo99}
F.~Dalfovo, S.~Giorgini, L.~P. Pitaevskii, and S.~Stringari, ``Theory of
  bose-einstein condensation in trapped gases,'' {\em Reviews of Modern
  Physics} {\bfseries 71} no.~3, (1999) 463.

\bibitem{Pavloff02}
N.~Pavloff, ``Breakdown of superfluidity of an atom laser past an obstacle,''
  {\em Phys. Rev. A} {\bfseries 66} (2002) 013610,
  \href{http://arxiv.org/abs/cond-mat/0206271}{{\ttfamily
  arXiv:cond-mat/0206271 [cond-mat]}}.

\bibitem{Cornish00}
S.~L. Cornish, N.~R. Claussen, J.~L. Roberts, E.~A. Cornell, and C.~E. Wieman,
  ``{Stable 85 Rb Bose-Einstein condensates with widely tunable
  interactions},'' {\em Phys. Rev. Lett.} {\bfseries 85} no.~9, (2000) 1795,
  \href{http://arxiv.org/abs/0004290}{{\ttfamily arXiv:0004290 [cond-mat]}}.

\bibitem{Unruh95}
W.~Unruh, ``{Sonic analog of black holes and the effects of high frequencies on
  black hole evaporation},''
\href{http://dx.doi.org/10.1103/PhysRevD.51.2827}{{\em Phys. Rev.} {\bfseries D
  51} (1995) 2827--2838}.

\bibitem{Corley96}
S.~Corley and T.~Jacobson, ``{Hawking spectrum and high frequency
  dispersion},'' \href{http://dx.doi.org/10.1103/PhysRevD.54.1568}{{\em Phys.
  Rev.} {\bfseries D 54} (1996) 1568--1586},
\href{http://arxiv.org/abs/hep-th/9601073}{{\ttfamily arXiv:hep-th/9601073
  [hep-th]}}.

\bibitem{Brout95}
R.~Brout, S.~Massar, R.~Parentani, and P.~Spindel, ``{Hawking radiation without
  transPlanckian frequencies},''
  \href{http://dx.doi.org/10.1103/PhysRevD.52.4559}{{\em Phys. Rev.} {\bfseries
  D 52} (1995) 4559--4568},
\href{http://arxiv.org/abs/hep-th/9506121}{{\ttfamily arXiv:hep-th/9506121
  [hep-th]}}.

\bibitem{Himemoto00}
Y.~Himemoto and T.~Tanaka, ``{A Generalization of the model of Hawking
  radiation with modified high frequency dispersion relation},''
  \href{http://dx.doi.org/10.1103/PhysRevD.61.064004}{{\em Phys. Rev.}
  {\bfseries D 61} (2000) 064004},
\href{http://arxiv.org/abs/gr-qc/9904076}{{\ttfamily arXiv:gr-qc/9904076
  [gr-qc]}}.

\bibitem{Unruh04}
W.~G. Unruh and R.~Schutzhold, ``{On the universality of the Hawking effect},''
  \href{http://dx.doi.org/10.1103/PhysRevD.71.024028}{{\em Phys. Rev.}
  {\bfseries D 71} (2005) 024028},
\href{http://arxiv.org/abs/gr-qc/0408009}{{\ttfamily arXiv:gr-qc/0408009
  [gr-qc]}}.

\bibitem{Coutant11}
A.~Coutant, R.~Parentani, and S.~Finazzi, ``{Black hole radiation with short
  distance dispersion, an analytical S-matrix approach},''
  \href{http://dx.doi.org/10.1103/PhysRevD.85.024021}{{\em Phys. Rev.}
  {\bfseries D 85} (2012) 024021},
\href{http://arxiv.org/abs/1108.1821}{{\ttfamily arXiv:1108.1821 [hep-th]}}.

\bibitem{Coutant14b}
A.~Coutant and R.~Parentani, ``{Hawking radiation with dispersion: The
  broadened horizon paradigm},''
  \href{http://dx.doi.org/10.1103/PhysRevD.90.121501}{{\em Phys. Rev.}
  {\bfseries D 90} (2014) 121501},
\href{http://arxiv.org/abs/1402.2514}{{\ttfamily arXiv:1402.2514 [gr-qc]}}.

\bibitem{Philbin16}
T.~Philbin, ``{An exact solution for the Hawking effect in a dispersive
  fluid},'' \href{http://dx.doi.org/10.1103/PhysRevD.94.064053}{{\em Phys.
  Rev.} {\bfseries D 94} no.~6, (2016) 064053},
\href{http://arxiv.org/abs/1607.03743}{{\ttfamily arXiv:1607.03743 [gr-qc]}}.

\bibitem{Parentani10}
R.~Parentani, ``{From vacuum fluctuations across an event horizon to long
  distance correlations},''
  \href{http://dx.doi.org/10.1103/PhysRevD.82.025008}{{\em Phys. Rev.}
  {\bfseries D 82} (2010) 025008},
\href{http://arxiv.org/abs/1003.3625}{{\ttfamily arXiv:1003.3625 [gr-qc]}}.

\bibitem{Schutzhold10}
R.~Schutzhold and W.~G. Unruh, ``{On Quantum Correlations across the Black Hole
  Horizon},'' \href{http://dx.doi.org/10.1103/PhysRevD.81.124033}{{\em Phys.
  Rev.} {\bfseries D 81} (2010) 124033},
\href{http://arxiv.org/abs/1002.1844}{{\ttfamily arXiv:1002.1844 [gr-qc]}}.

\bibitem{Starobinskii73}
A.~Starobinskii, ``Amplification of waves during reflection from a rotating
  black hole,'' {\em Zh. Eksp. Teor. Fiz} {\bfseries 64} (1973) 48.

\bibitem{Barcelo05}
C.~Barcelo, S.~Liberati, and M.~Visser, ``{Analogue gravity},'' {\em Living
  Rev. Rel.} {\bfseries 8} (2005) 12,
\href{http://arxiv.org/abs/gr-qc/0505065}{{\ttfamily arXiv:gr-qc/0505065
  [gr-qc]}}.

\bibitem{Busch14}
X.~Busch and R.~Parentani, ``{Quantum entanglement in analogue Hawking
  radiation: When is the final state nonseparable?},''
  \href{http://dx.doi.org/10.1103/PhysRevD.89.105024}{{\em Phys. Rev.}
  {\bfseries D 89} no.~10, (2014) 105024},
\href{http://arxiv.org/abs/1403.3262}{{\ttfamily arXiv:1403.3262 [gr-qc]}}.

\bibitem{Werner89}
R.~F. Werner, ``{Quantum states with Einstein-Podolsky-Rosen correlations
  admitting a hidden-variable model},'' {\em Phys. Rev.} {\bfseries A 40}
  no.~8, (1989) 4277.

\bibitem{Finke16}
A.~Finke, P.~Jain, and S.~Weinfurtner, ``{On the observation of nonclassical
  excitations in Bose--Einstein condensates},''
  \href{http://dx.doi.org/10.1088/1367-2630/18/11/113017}{{\em New J. Phys.}
  {\bfseries 18} no.~11, (2016) 113017},
\href{http://arxiv.org/abs/1601.06766}{{\ttfamily arXiv:1601.06766
  [quant-ph]}}.

\bibitem{Michel16}
F.~Michel, J.-F. Coupechoux, and R.~Parentani, ``{Phonon spectrum and
  correlations in a transonic flow of an atomic Bose gas},''
  \href{http://dx.doi.org/10.1103/PhysRevD.94.084027}{{\em Phys. Rev.}
  {\bfseries D 94} no.~8, (2016) 084027},
\href{http://arxiv.org/abs/1605.09752}{{\ttfamily arXiv:1605.09752
  [cond-mat.quant-gas]}}.

\bibitem{Fabbri10}
A.~Fabbri and C.~Mayoral, ``{Step-like discontinuities in Bose-Einstein
  condensates and Hawking radiation: the hydrodynamic limit},''
  \href{http://dx.doi.org/10.1103/PhysRevD.83.124016}{{\em Phys. Rev.}
  {\bfseries D83} (2011) 124016},
\href{http://arxiv.org/abs/1004.4876}{{\ttfamily arXiv:1004.4876 [gr-qc]}}.

\bibitem{Olver}
F.~Olver, {\em Asymptotics and special functions}, vol.~15.
\newblock Academic Press New York, 1974.

\bibitem{DLMF}
F.~W. Olver, D.~W. Lozier, R.~F. Boisvert, and C.~W. Clark, ``Digital library
  of mathematical functions,'' {\em National Institute of Standards and
  Technology from http://dlmf. nist. gov/(release date 2011-07-01), Washington,
  DC} (2010) .

\bibitem{Berry72}
M.~V. Berry and K.~Mount, ``Semiclassical approximations in wave mechanics,''
  {\em Reports on Progress in Physics} {\bfseries 35} no.~1, (1972) 315.

\bibitem{Christensen77}
S.~M. Christensen and S.~A. Fulling, ``{Trace Anomalies and the Hawking
  Effect},''
\href{http://dx.doi.org/10.1103/PhysRevD.15.2088}{{\em Phys. Rev.} {\bfseries D
  15} (1977) 2088--2104}.

\end{thebibliography}\endgroup

\end{document}